\documentclass[prx,aps,amssymb,twocolumn,superscriptaddress]{revtex4-1}
\usepackage{amsmath}
\usepackage{amssymb}
\usepackage{amsthm}
\usepackage{amsfonts}
\usepackage{listings}
\usepackage{longtable}
\usepackage{enumerate}
\usepackage{latexsym}
\usepackage{color}
\usepackage{setspace} 
\usepackage{blindtext}
\usepackage{dsfont}
\usepackage{mathrsfs}
\usepackage{array,etoolbox}
\preto\tabular{\setcounter{magicrownumbers}{0}}
\newcounter{magicrownumbers}

\usepackage{bm}
\usepackage{hyperref}

\newcommand{\beginsupplement}{
        \setcounter{table}{0}
        \renewcommand{\thetable}{S\arabic{table}}
        \setcounter{figure}{0}
        \renewcommand{\thefigure}{S\arabic{figure}}
        \setcounter{equation}{0}
        \renewcommand{\theequation}{S\arabic{equation}}
     }

\usepackage{psfrag}

\usepackage{bm}
\usepackage{graphicx}
\usepackage{subfigure}
\RequirePackage[normalem]{ulem} %DIF PREAMBLE
\RequirePackage{color}\definecolor{RED}{rgb}{1,0,0}\definecolor{BLUE}{rgb}{0,0,1} %DIF PREAMBLE
\newcommand{\beq}{\begin{equation}}
\newcommand{\eneq}{\end{equation}}
\input{epsf}

\begin{document}

\tolerance 10000

\newcommand{\vk}{{\bf k}}

%\draft

\title{Topology-Bounded Superfluid Weight In Twisted Bilayer Graphene }

\author{Fang Xie}
\thanks{These two authors contributed equally.}
\author{Zhida Song}
\thanks{These two authors contributed equally.}
\affiliation{Department of Physics, Princeton University, Princeton, New Jersey 08544, USA}
\author{Biao Lian}
\affiliation{Princeton Center for Theoretical Science, Princeton University, Princeton, New Jersey 08544, USA}
\author{B. Andrei Bernevig}
\email{bernevig@princeton.edu}
\affiliation{Department of Physics, Princeton University, Princeton, New Jersey 08544, USA}
\affiliation{Physics Department, Freie Universitat Berlin, Arnimallee 14, 14195 Berlin, Germany}
\affiliation{Max Planck Institute of Microstructure Physics, 
06120 Halle, Germany
}
 
\date{\today}
%\pacs{03.67.Mn, 05.30.Pr, 73.43.-f}

\begin{abstract}
While regular flat bands are good for enhancing the density of states and hence the gap, they are detrimental to the superfluid weight. We show that the predicted nontrivial topology of the two lowest flat bands of twisted bilayer graphene plays an important role in the enhancement of the superfluid weight and hence of superconductivity. We derive the superfluid weight (phase stiffness) of the TBLG  superconducting flat bands with a uniform pairing, and show that it can be expressed as an integral of the Fubini-Study metric of the flat bands.  This mirrors results \cite{Peotta2015} already obtained for nonzero Chern number bands even though the TBLG flat bands have zero Chern number.
We further show the metric integral is lower bounded by the topological $C_{2z}T$ Wilson loop winding number of the TBLG flat bands, which renders the superfluid weight has a topological lower bound proportional to the pairing gap. In contrast, trivial flat bands have a zero superfluid weight. The superfluid weight is crucial in determining the BKT transition temperature of the superconductor. Based on the transition temperature measured in TBLG experiments, we estimate the topological contribution of the superfluid weight in TBLG.\end{abstract}

\maketitle

The recently discovered superconducting phase in twisted bilayer graphene has received extensive attention \cite{morell2010,bistritzer2011,cao2018,yankowitz2018,huangs2018,
yuan2018,po2018,xu2018,roy2018,dodaro2018,wu2018, isobe2018,huang2018,you2018,wux2018,kang2018,kennes2018,guinea2018,peltonen2018,fidrysiak2018,zou2018,gonz2018,su2018,guo2018,songz2018,po2018b,ahn2018,hejazi2018,laksono2018,tarnopolsky2018,venderbos2018,chenl2018,stauber2018,choi2018,jian2018a,kozii2018,kang2018b,wux2018b,wuf2018b,liuj2018,Lian2018,polshyn2019}. The topology of the lowest two bands (per spin and valley) of twisted bilayer graphene (TBLG) is currently under debate \cite{zou2018,songz2018,po2018b,ahn2018,po2018f}. Although theoretical models suggest a nontrivial topological number of these bands, the experimentally measurable effects through which one could prove or falsify the predicted nontrivial topology are scarce. Currently, one viable experimentally observable effect \cite{lian2018landau}, predicts that the single-particle magnetic field spectrum of a topologically nontrivial band can cross the single-particle gap, in stark contrast to conventional knowledge and to the in-field spectrum of trivial bands. We here present another effect of a set of topologically nontrivial bands observable at zero field (of the kind present in TBLG) that appears when these bands become superconducting. We show that the superfluid weight in the superconducting state is the sum of two terms: a conventional term, which vanishes when the bands are perfectly flat, and a topological term, bounded from below by the Wilson loop winding number of the $C_{2z}T$ protected topology in TBLG. 

This letter is organized as follows. First, we show that by assuming perfectly flat bands and $s$ wave pairing, the superfluid weight can be written as the integral of Fubini-Study metric over the Brillouin zone (BZ), and show that it is lower-bounded by the Wilson loop winding. Secondly, by applying this result to TBLG, we estimate the topological contribution of superfluid weight and explain the relatively high transition temperature.

The two characterizing features of superconductors are the zero DC resistance and Meissner effect. Both of these properties are captured by the celebrated London equation \cite{tinkham2004introduction}. It tells us that the electric current in a superconductor $\mathbf{j}$ is proportional to the gauge potential $\mathbf{A}$ under Coulomb gauge:
\begin{equation}\label{eqn:londonequation}
	j_i = - [D_s]_{ij}A_j\,,
\end{equation} in which the coefficient $[D_s]_{ij}$ is called the superfluid weight and it is a tensor in general. Some spacial symmetry, such as $C_{3z}$, requires it to be isotropic in 2D. In some works, such as Ref. \cite{corson1999vanishing}, it is called ``phase stiffness'' (describing energy susceptibility with respect to phase twists). The London equation has two kinds of consequences. One is the perfect diamagnetism feature of superconductor, and the other one is the frequency dependence of AC conductance. In 2D, the superfluid weight is also related with the transition temperature. The phase coherence will disappear at a temperature $T= T_c$ given by  $\frac{\hbar^2D_s(T_c)}{e^2k_BT_c}=\frac{8}{\pi}$ (known as Berezinskii-Kosterlitz-Thouless (BKT) transition \cite{Berezinsky1970,Kosterlitz1973}), because of the creation of vortex-antivortex pairs. Usually $D_s(T)$ decreases with increasing temperature, so the transition temperature $T_c$ is always lower than $\frac{\pi\hbar^2D_s(0)}{8e^2k_B}$. Thus a small $D_s$ at zero temperature leads to a low transition temperature.

Two methods, using a static magnetic field or an alternating electric field, are used to measure the superfluid weight (phase stiffness) directly. In 3D, $D_s$ can be obtained by measuring the penetration depth $\lambda_L=\sqrt{1/\mu_0 D_s}$ of a static magnetic field into the superconductor. This method, however, fails in 2D, as the penetration depth is no longer a simple function of $D_s$. Instead, one can measure the AC conductance under an alternating electric field of frequency $\omega$. By taking time derivative on both sides of Eq. (\ref{eqn:londonequation}), we obtain the frequency dependent conductance, and it is inversely proportional to the frequency as $\sigma(\omega) = \frac{iD_s}{\omega}$. Therefore the superfluid weight will be $D_s = -i\omega \sigma(\omega)$. In experiments, $\sigma(\omega)$ can be measured by the time-domain transmission spectroscopy without any contact with the sample \cite{gruner1998}. For example, at zero temperature, the stiffness temperature of $\rm Bi_2Sr_2CaCu_2O_{8+\delta}$ is measured to be $T_\theta = 55\rm\,K$, and the corresponding superfluid weight is $D_s = e^2 k_B T_\theta/\hbar^2 = 1.8 \times 10^{9}\,\rm H^{-1}$ \cite{corson1999vanishing}. As another example, the superfluid weight of MoGe thin film is measured to be $D_s = 5 \times 10^8 \,\rm H^{-1}$ \cite{Misra2013}.

In Landau-Ginzburg (LG) theory of conventional superconductivity, the superfluid weight is given by $D_s\approx e^2n_s/m^*$ in which $m^*$ is the band effective mass and $n_s$ is the superfluid density \cite{scanlapino1992superfluid,tinkham2004introduction}. At zero temperature, all the electrons have contribution to superconducting transport, which means $n_s(T=0)$ is equal to the total electron density, and $n_s(T)$ usually decreses with increasing temperature. If the band is exactly flat, the band mass will become infinity, and the LG theory tells us the superfluid weight can be zero even when Cooper pairing happens. We use the Bistritzer-MacDonald model to estimate the bandwidth and the conventional contribution of superfluid weight in TBLG. Around the magic angle, the flat band mostly lie in an energy range $|\varepsilon| < W \approx 0.5\,\rm meV$. Hence the effective mass is approximately $m^* \approx \hbar^2K_M^2/2W$, where $K_M$ is the distance between $\Gamma$ and $K$ in Moir\'e BZ. Thus the conventional superfluid weight is $[D_s]_{\rm tri} \approx e^2n_s/m^* \approx 2e^2WN/\hbar^2\Omega_c K_M^2 = 3\sqrt{3}e^2WN/4\pi^2\hbar^2$, where $\Omega_c$ is the area of the Moir\'e unit cell, and $N$ is the number of electrons per Moir\'e unit cell. Here we assume that the superfluid density is given by the total electron density, which is the upper limit of $n_s$. If we consider the case with filling $\nu = 1/4$ or equivalently $N=2$, the value of superfluid weight will be $[D_s]_{\rm tri} \approx 5\times 10^{7}\,\rm H^{-1}$, and the corresponding BKT transition temperature will not be higher than $0.6\,\rm K$. However, LG theory is valid only when the band is trivial, as the spreading of its Wannier function has a nonzero lower bound, therefore the estimation based on LG theory in this paragraph is not enough \cite{Peotta2015,tovmasyan2016,altman2010bose}. As a result, we show that even in the exactly flat band limit, the Cooper pairing may acquire nonlocal phase correlations and thus a nonzero superfluid weight, which - as we show - gives rise to a higher transition temperature.

To obtain the contribution of nontrivial band topology to the superfluid weight, we consider a mean-field Bogoliubov-de-Gennes (BdG) Hamiltonian of a superconductor:
\begin{align}
	H_{\rm BdG} =&\frac{1}{2}\sum_{\mathbf{k}}\Psi_{\mathbf{k}}^\dagger\left(
		\begin{array}{cc}
			\mathcal{H}(\mathbf{k})-\mu & \Delta(\mathbf{k}) \\ \Delta^\dagger(\mathbf{k}) & -\mathcal{H}^*(-\mathbf{k}) + \mu
		\end{array}
	\right)\Psi_{\mathbf{k}} \nonumber\\
	&+ \frac12 \sum_\mathbf{k}{\rm Tr\,}(\mathcal{H}(\mathbf{k})-\mu)\,. \label{eqn:bdg} 
\end{align}
We use $\Omega_0$ to denote the ground state energy of $H_{\rm BdG}$, which is also the free energy at zero temperature. We substitute $\mathbf{k}\rightarrow \mathbf{k}-e\mathbf{A}$ by Peierls substitution \cite{Luttinger1951} when a non-zero uniform gauge potential $\mathbf{A}$ is turned on, and the free energy becomes a function of $\mathbf{A}$. We can then expand $\Omega(\mathbf{A})$ to the second order of $A_i$ and obtain $\Omega(\mathbf{A})\approx \Omega_0 + \frac12 V[D_s]_{ij}A_iA_j$, where $V$ is the area of the sample, and the second order coefficient $[D_s]_{ij}$ is the superfluid weight. The first order derivative $\partial_{\mathbf{A}}\Omega(\mathbf{A})$ gives us the electric current, which agrees with the London equation shown in Eq. (\ref{eqn:londonequation}). 

The free energy $\Omega(\mathbf{A})$ and thus the superfluid weight can be derived from the BdG Hamiltonian. The general expression of the superfluid weight is the summation of three terms given by Eqs. (\ref{eqn:swterm1}), (\ref{eqn:swterm2}) and (\ref{eqn:swterm3}) in the supplementary material (SM) Sec. \ref{sec:mfsw}. The first term in Eq. (\ref{eqn:swterm1}) corresponds to the Landau-Ginzburg contribution, while Eqs. (\ref{eqn:swterm2}) and (\ref{eqn:swterm3}) are additional contributions due to the $\mathbf{k}$ dependence of the flat band Bloch wave functions. When the bands are flat, the conventional contribution vanishes, but the wave function contributions Eqs. (\ref{eqn:swterm2}) and (\ref{eqn:swterm3}) can be nonzero, and are related to the band topology as we show below.

Before we start our discussion about TBLG, we briefly review the superconductivity in the spin Chern insulator with exactly flat bands studied in Ref. \cite{Peotta2015}. In this model, $\mathcal{H}(\mathbf{k})$ has both spinful time reversal symmetry and $s_z$ conservation, which allows one to define a spin Chern number $C$. The order parameter $\Delta(\mathbf{k}) = is_y\Delta$ (in which $s_y$ is the $y$ direction spin Pauli matrix) is momentum independent, and one can show that the superfluid weight given by sum of Eqs. (\ref{eqn:swterm1}), (\ref{eqn:swterm2}) and (\ref{eqn:swterm3}) can be reduced to the following integral in the BZ:
\begin{equation}\label{eqn:peottaresult}
	[D_s]_{ij} = \frac{8e^2 \Delta}{\hbar^2} \sqrt{\nu(1-\nu)}\int\frac{d^2k}{(2\pi)^2}g_{ij}(\mathbf{k})\,,
\end{equation}
where $\nu$ is the filling ratio of the spinful flat bands, and $g_{ij}(\mathbf{k})$ is the Fubini-Study metric evaluated from the Bloch wave function of the spin $\uparrow$ flat band:
\begin{align}\label{eqn:abelianmetric}
	g_{ij}(\mathbf{k}) = & \frac12 \left(\partial_{k_i}u^\dagger(\mathbf{k})\partial_{k_j}u(\mathbf{k}) + \partial_{k_j}u^\dagger(\mathbf{k})\partial_{k_i}u(\mathbf{k})\right)\nonumber\\
	& + u^\dagger(\mathbf{k})\partial_{k_i}u(\mathbf{k})u^\dagger(\mathbf{k})\partial_{k_j}u(\mathbf{k})\,,
\end{align}
where $u(\mathbf{k})$ is the Bloch wave function at momentum $\mathbf{k}$ of the spin up flat band. 
%The spin down wave function at $\mathbf{k}$ is related to $u(-\mathbf{k})$ by spinful time reversal symmetry, and that's why the superfluid weight can be easily expressed by using the spin up wave function only. 
This result is derived in the exact flat band limit, so the contribution from the band dispersion (Eq. (\ref{eqn:swterm1})) disappears. Thus we discover that LG theory prediction of superfluid weight is not enough even when the flat band is trivial, because we have models which has trivial bands and $\mathbf{k}$ dependent wave function $u(\mathbf{k})$, such as the ``topological'' phase of SSH model \footnote{The ``topological'' phase of SSH model is trivial because it has maximally localized Wannier state, although its Wannier function localization functional is bounded by the winding number from below}. A nonzero spin Chern number further gives a nonzero lower bound of the Fubini-Study metric, which will be discussed in the following paragraph.

The Fubini-Study metric defines a distance on the BZ torus: two momentum points are close to each other if their wave functions have a large overlap \cite{cheng2010quantum}. The integral of ${\rm tr}\,g = g_{xx} + g_{yy}$ also corresponds to the gauge invariant part of the ``Wannier function localization functional'', or the spread functional, which has been studied in detail in previous research \cite{Vanderbelt1997Wannier,Vanderbilt2012Wannier}. The metric is also related to Berry curvature through the quantum geometric tensor defined by $\mathfrak{G}_{ij} = \partial_{k_i}u^\dagger(\mathbf{k})(1-u(\mathbf{k})u^\dagger(\mathbf{k}))\partial_{k_j}u(\mathbf{k})$. The real part of $\mathfrak{G}_{ij}$ is the metric $g_{ij}$ and the imaginary part is the Berry curvature. One of the most important properties of $\mathfrak{G}_{ij}$ is its \emph{positive definiteness}. It can be shown that for arbitrary complex vectors $\{c_i\}$, the inequality $\sum_{ij}c^\dagger_i\mathfrak{G}_{ij}c_j \geq 0$ always holds \footnote{See SM Sec. \ref{sec:metric}}. If we choose $c_x = 1$ and $c_y = i$, we will find ${\rm tr}\, g = g_{xx} + g_{yy} \geq -\mathcal{F}_{xy}$; similarly we choose $c_x = 1$ and $c_y = -i$, and we will obtain ${\rm tr}\,g\geq \mathcal{F}_{xy}$. Therefore we prove that the metric is bounded by the absolute value of curvature ${\rm tr}\,g \geq|\mathcal{F}_{xy}|$. From the expression of $D_s$ one can easily notice ${\rm tr}\,D_s$ is bounded by the spin Chern number $C$. More details of the quantum metric is discussed in supplementary material Sec. \ref{sec:metric}. In TBLG, the (spin) Chern number is zero, and the system is multi-band, likely with more complicated pairing symmetry. Hence a new bound/limit (if any exists) for the superfluid weight must be obtained. 

We first generalize the result of Ref. \cite{Peotta2015} to multi-band systems with a more realistic pairing. The free fermion Hamiltonian $\mathcal{H}(\mathbf{k})$ is assumed to be invariant under spinful time reversal transformation, which is represented by $\mathcal{T} = U_T \mathcal{K}$, where $U_T$ is a real unitary matrix and $\mathcal{K}$ is complex conjugation operator. We do not (any longer) assume momentum independent pairing. $\mathcal{H}(\mathbf{k})$ is diagonalized by $U(\mathbf{k})$ as $\varepsilon_\mathbf{k} = U^\dagger(\mathbf{k})\mathcal{H}(\mathbf{k})U(\mathbf{k})$ where $\varepsilon_\mathbf{k}$ is a diagonal matrix, and we assume that it has $N_F$ flat bands at same energy. We also assume the band gap between these flat bands and any other bands is larger than the bandwidth of flat bands, and the interaction between the electrons. In the following discussion we use a time reversal symmetric pairing between Kramers pairs as follows: 
\begin{equation}\label{eqn:pairing}
\Delta(\mathbf{k}) = \left[\Delta_1 \tilde{U}_\mathbf{k}\tilde{U}^\dagger_\mathbf{k}+ \Delta_2\left(\mathds{1}-\tilde{U}_\mathbf{k}\tilde{U}^\dagger_\mathbf{k}\right)\right]U_T\,,
\end{equation} 
in which $\Delta_{1,2}\in \mathbb{R}$ because of time reversal symmetry. Here $\tilde{U}_{\mathbf{k}} = (u_{1}(\mathbf{k}),u_2(\mathbf{k}),\cdots,u_{N_F}(\mathbf{k}))$ is the projection of $U(\mathbf{k})$ into these $N_F$ flat bands and $u_i(\mathbf{k})$ are the eigenvectors of matrix $\mathcal{H}(\mathbf{k})$. This ansatz implies that $s$ wave pairing happens between Kramers pairs and the pairing strength in the flat bands and in all other bands are given by $\Delta_1$ and $\Delta_2$, respectively. Therefore we have the following three important assumptions in total: 1) the free fermion Hamiltonian with time reversal symmetry has $N_F$ flat bands at the same energy near the Fermi level; 2) there is a large band gap between flat bands and other bands; 3) the pairing order parameter satisfies Eq. (\ref{eqn:pairing}). Because of this large band gap between flat bands and other bands, we can project the BdG Hamiltonian into the flat bands of the free fermion model and then derive the superfluid weight. The result is given by
\begin{equation}\label{eqn:sw}
 	[D_s]_{ij} = \frac{2e^2|\Delta_1|}{\hbar^2}\left(1 + \frac{\Delta_2}{\Delta_1}\right)\sqrt{\nu(1-\nu)}\int\frac{d^2k}{(2\pi)^2}g_{ij}(\mathbf{k})
  \end{equation}
\begin{align}
 	g_{ij}(\mathbf{k})=&{\rm Tr}\left[\frac{1}{2}\left(\partial_{k_i}\tilde{U}^\dagger_\mathbf{k}\partial_{k_j}\tilde{U}_\mathbf{k}+\partial_{k_j}\tilde{U}^\dagger_\mathbf{k}\partial_{k_i}\tilde{U}_\mathbf{k}\right)\right. \nonumber\\
 	& \left.+ \left( \tilde{U}^\dagger_\mathbf{k}\partial_{k_i}\tilde{U}_\mathbf{k}\tilde{U}^\dagger_\mathbf{k}\partial_{k_j}\tilde{U}_\mathbf{k} \right)\right]\label{eqn:nonabelianmetric}\,,
\end{align} 
in which ${\rm Tr}(X)=\sum_{n=1}^{N_F}(X_{nn})$ stands for the trace over all the flat band indices. Eq. (\ref{eqn:nonabelianmetric}) is the the generalization of Fubini-Study metric in Eq. (\ref{eqn:abelianmetric}) to multi-band systems, which is still positive definite \footnote{See SM Sec. \ref{sec:fbsw}.}. The result in Eq. (\ref{eqn:peottaresult}) derived in Ref. \cite{Peotta2015} is a special case of our result in Eq. (\ref{eqn:sw}) when the time reversal transformation is represented by $U_T = is_y$, the spin $z$ component is conserved, and $\Delta = \Delta_1 = \Delta_2$.

We also notice that if we assume $\frac{\Delta_2}{\Delta_1} \leq -1$, the superfluid weight will become zero, or even a negative number. A negative superfluid weight is unphysical, denoting that the BCS wave function of such a pairing is not a stable ground state. This also indicates that if there is no constraint on the order parameter $\Delta(\mathbf{k})$, the superfluid weight will not be bounded. However we expect a weaker pairing strength in the bands which are farther away from the Fermi level, or $|\Delta_2| < |\Delta_1|$. If the pairing in higher bands are much stronger than pairing in the flat bands - a physically impossible situation -, our projection of the BdG Hamiltonian into the flat bands may also become invalid. Hence we later set $\Delta_2 = 0$ in order to estimate the topological contribution of $D_s$.

Now we apply Eq. (\ref{eqn:sw}) to TBLG, and show that the fragile topology of TBLG flat bands yields a finite lower bound of the superfluid weight although it has zero spin Chern number. The Bistritzer-MacDonald (BM) model \cite{bistritzer2011} has $C_{2z} T$, $C_{2x}$ and $C_{3z}$ symmetries, in which $T$ stands for the spinless time reversal transformation. If all the spins and valleys are considered here, we will have well-defined time reversal symmetry, although BM model itself does not. The $C_{2z}T$ symmetry is crucial for the flat bands' topology \cite{songz2018,ahn2018,zou2018,po2018f}. Because of this symmetry, the two eigenvalues of the non-Abelian Wilson loop have to be complex conjugation to each other \cite{songz2018,ahn2018}. A winding number can be defined by Wilson loop eigenvalues. $C_{2z} T$ symmetry gives a constraint not only to the Wilson loop but also to the Berry connection and Berry curvature. It can be shown \cite{songz2018,ahn2018} that the non-Abelian Berry connection and curvature of the two flat bands can always be written as $\mathbf{A}(\mathbf{k}) = -\mathbf{a}(\mathbf{k})\sigma_2$ and $\mathcal{F}_{xy}(\mathbf{k}) = -f_{xy}(\mathbf{k})\sigma_2$ under a proper \emph{local} gauge choice on a patch in the Brillouin zone (although a \emph{global} gauge choice which satisfies this condition might not exist \footnote{See SM Sec. \ref{sec:lowerbound}. The transformation $g_\mathbf{k}$ can transform $\mathbf{A}(\mathbf{k})$ into $-\mathbf{a}(\mathbf{k})\sigma_2$. However we notice it is not a single value function of $\mathbf{k}$. It changes the sign when it goes around a Dirac point.}.) In SM Sec. \ref{sec:lowerbound}, we prove that the Wilson loop winding number \cite{songz2018}, or the ``Euler class'' in Ref. \cite{ahn2018} denoted by $e_2$, of the two topological bands, which is an integer (if there are more than two bands, the topological classification will be $\mathbb{Z}_2$ instead of $\mathbb{Z}$), is given by the integral of $f_{xy}$ over the whole BZ (with Dirac points removed from the integral area):
\begin{equation}
	e_2 = \frac{1}{2\pi}\int d^2k\,f_{xy}\,.
\end{equation}
The wave function of the two flat bands in TBLG  (per spin per valley - which are good quantum numbers for small twist angles) also can be used to define the positive-definite non-Abelian quantum geometric tensor $\mathfrak{G}_{ij}$ (which is a $2\times 2$ complex matrix) and Fubini-Study metric $g_{ij} = \frac12{\rm Tr}\,(\mathfrak{G}_{ij} + \mathfrak{G}_{ij}^\dagger)$. For arbitrary complex vectors $c_i\in\mathbb{C}^2$, the inequality $\sum_{ij}c^\dagger_i \mathfrak{G}_{ij} c_j \geq 0$ always holds \footnote{See SM Sec. \ref{sec:metric}.}. By choosing vectors $c_x$ and $c_y$ properly \footnote{We show how we can choose proper $c_x$ and $c_y$ in SM Sec. \ref{sec:lowerbound}.}, we find that the metric is bounded by the ``Abelian part'' $f_{xy}$ of the non-Abelian Berry curvature $\mathcal{F}_{xy}$: ${\rm tr}\,g \geq 2|f_{xy}|$. The derivation of band topology and metric of the TBLG can be found in SM Sec. \ref{sec:lowerbound}.

In small angle TBLG, all bands are 4-fold degenerate with respect to spin $\uparrow, \downarrow$ and with respect to original graphene valley $K$, $K'$. The order parameter ansatz in Eq. (\ref{eqn:pairing}) corresponds to the pairing between opposite spins and valleys, because time reversal transformation in TBLG will flip both spin $\uparrow,\downarrow$ and valley $K,K'$. If we assume no pairing in higher bands ($\Delta_1 = \Delta, \Delta_2=0$), which is physically reasonable, the superfluid weight in the exact flat band limit will be the same as the expression shown in Eq. (\ref{eqn:peottaresult}). However, both the pairing strength and metric have different meanings. $g_{ij}(\mathbf{k})$ stands for the Fubini-Study metric derived from the wave functions of two flat bands (of with spin $\uparrow$ and valley $K$, all other degenerate spin and valley bands have identical contribution to the superfluid weight). Also $\Delta_1$ is no longer the pairing strength in all the bands but only in these flat bands. Because of the two band winding number protected by $C_{2z}T$ symmetry, we have a new lower bound. The inequality ${\rm tr}\,g > 2|f_{xy}|$ naturally leads to the lower bound of the trace of the superfluid weight ${\rm tr}\,D_s \geq \frac{8e^2\Delta_1}{\pi \hbar^2}\sqrt{\nu(1-\nu)}$ even though the (spin) Chern number here vanishes. Here we used the fact that the winding number of TBLG flat bands is $e_2 = 1$. As we mentioned earlier, $C_{3z}$ symmetry requires the superfluid weight tensor to be isotropic, therefore $D_s = \frac12 {\rm tr}\,[D_s]_{ij}$. Then we can use the following equation to estimate the topological contribution of $D_s$:
\begin{equation}\label{eqn:topoDs}
 	D_s \approx \frac{4e^2\Delta_1}{\pi\hbar^2}\sqrt{\nu(1-\nu)}|e_2|.
\end{equation} 
From this equation, we find that a nonzero superfluid weight is possible even when the bands are exactly flat, as long as Cooper pairing gap is developed ($\Delta_1 \neq  0$).  The parameter $\Delta_1$ can be estimated from the measured $T_c$. The filling ratio $\nu$ is determined by the carrier density. Below we use experimental data to estimate the value of superfluid weight at zero temperature.

For flat bands with attractive interaction between Kramers pairs, the mean field BCS wave function is a good approximation to the ground state (SM Sec. \ref{sec:validity}). To see this, we can perform a partial particle-hole transformation $c_{\mathbf{k},\uparrow} \rightarrow c_{\mathbf{k}\,\uparrow}\,,c_{\mathbf{k}\downarrow}\rightarrow c^\dagger_{-\mathbf{k}\downarrow}$, under which the attractive interaction becomes repulsive, and the BCS wave function becomes a ferromagnetic state. Since repulsive interaction favors ferromagnetic state according to the Hund's rule, we conclude that BCS wave function describes the ground state well.

\begin{figure}[!htp]
\centering
\includegraphics[width=8cm]{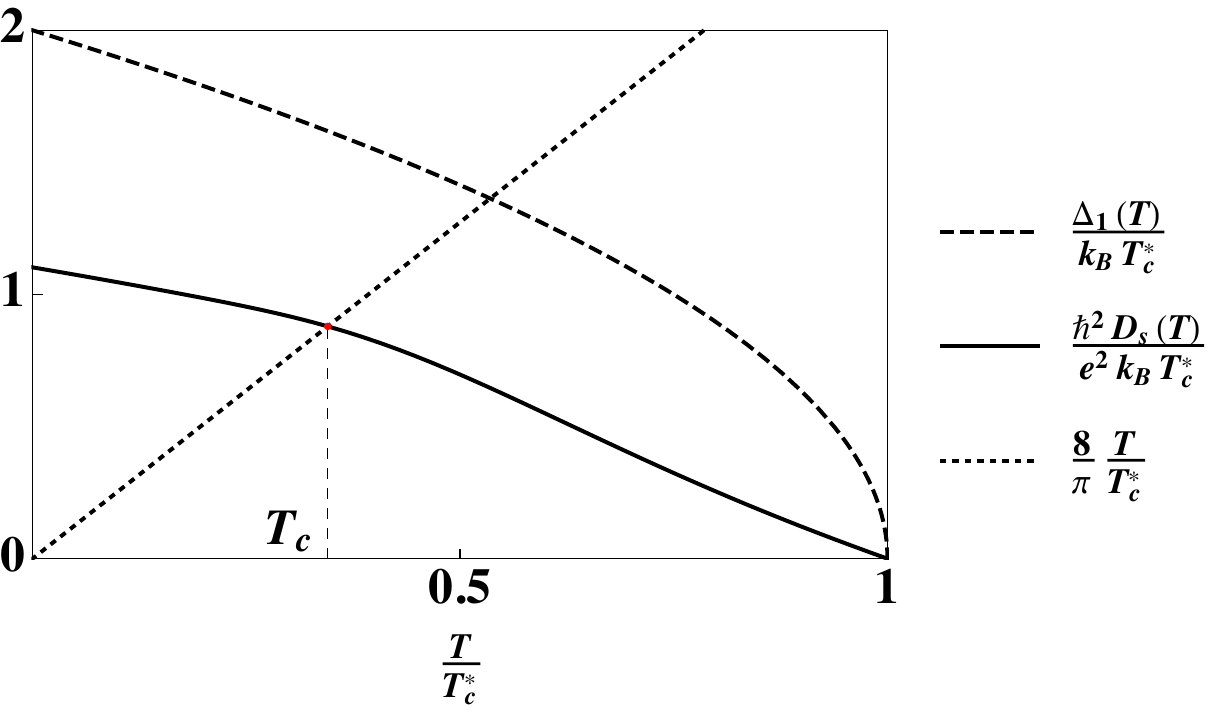}
\caption{The gap function $\Delta_1(T)$ and superfluid weight $D_s(T)$. In the flat band limit the BKT temperature is $T_c \approx 0.35 T_c^*$ when $\nu = 1/4$.}
\label{fig:KT}
\end{figure}

As previously mentioned, in 2D superconductors, the transition temperature $T_c$ measured in experiment is the Berezinskii-Kosterlitz-Thouless temperature (when phase coherence disappears) instead of the BCS mean field transition temperature $T_c^* \approx \Delta_1/2k_B$ (when Cooper pairing vanishes). The BKT transition temperature is given by the universal relation $\frac{\hbar^2 D_s(T_c)}{e^2k_BT_c} = \frac{8}{\pi}$ \cite{Kosterlitz1973}, and it is generically lower than the BCS mean field transition temperature $T_c^*$. To derive the BKT temperature of TBLG in the flat band limit, we can generalize the topological superfluid weight expression in Eq. (\ref{eqn:topoDs}) to finite temperatures, which however has no simple analytical expression. By assuming $\Delta_1(T) \approx 2k_BT_c^*(1-T/T_c^*)^{1/2}$ \cite{tinkham2004introduction}, we can numerically calculate the temperature dependence of the superfluid weight. As an example, we have plotted $D_s$ in FIG. \ref{fig:KT} as a function of $T_c/T_c^*$ for filling ratio $\nu=1/4$ (2 electrons per Moir\'e unit cell), from which we find $T_c/T_c^*=0.35$. See SM Sec. \ref{sec:ftbkt} for more detailed calculations.

We now estimate the TBLG topological superfluid weight $D_s$ at zero temperature. When the bandwidth is small (zero), one expects the superfluid weight to be dominated by the band topology contribution. The experimental transition temperature \cite{cao2018,yankowitz2018} is $T_c = 1.5 \rm\,K$, thus the order parameter can be estimated to be $\Delta_1 = 2k_B T_c^* \approx 0.74 \rm\,meV$. By using $\nu = 1/4$, the topological superfluid weight is $[D_s]_{\rm top} \approx \frac{4e^2\Delta_1}{\pi \hbar^2}\sqrt{\nu(1-\nu)} \approx 1.5 \times 10^8 \rm \, H^{-1}$. One notices that this is an order of magnitude smaller than the superfluid weight in conventional materials, such as BSCCO and MoGe \cite{kivelson1995,basov2011manifesto}. However in both cuprates and MoGe thin films, the typical carrier density is around $n \approx 10^{14}\,\rm cm^{-2}$ \cite{Uemura1989} two orders of magnitude larger than that in TBLG (where $n\approx 10^{12}\,\rm cm^{-2}$), hence such a superfluid weight is already large for TBLG.

Moving away from completely flat bands, we find that the conventional term in the superfluid weight mostly depends on the bandwidth $W$, while the topological term mostly depends on the pairing order parameter $\Delta_1$ (or the transition temperature $T_c$). In TBLG, the bandwidth $W$ and the transition temperature $T_c$ have a similar  magnitude, and  we expect the topological term to have an important contribution. In the strong pairing limit where $\Delta > W$, the superfluid weight will be underestimated if by only the conventional contribution \cite{stauber2018}.

Recently, superconductivity has been observed in other Moir\'e systems, including twisted double bilayer graphene, and multilayer graphene/boron nitride heterostructure \cite{zhangy2018,zhangy2018b,chenguorui2018,moriyama2019,chenguorui2019,chittari2019,shen2019,liux2019,endo2019}. In these systems, a displacement electric field is necessary for superconductivity. This yields gapped flat bands with a nonzero valley Chern number \cite{zhangy2018,chittari2019}. The Chern number can be larger than one, which could lead to a larger topological lower bound of superfluid weight. In twisted multilayer graphene, the bandwidth of flat bands is even smaller than that in TBLG. A higher transition temperature (larger $\Delta$) was also observed \cite{shen2019,liux2019}. We expect the topological lower bound plays a significant role in the superfluid weight of these systems. Moreover, multi-layer systems with $C_{2z}T$ symmetry can realize larger Euler $e_2$ invariants (higher Wilson loop windings), hence increasing the superfluid weight. 

In summary, we proved that fragile topology can yield a nonzero superfluid weight (phase stiffness) of a superconductor even when the bands are exactly flat. In TBLG, by assuming exact flat bands and uniform $s$ wave pairing, the superfluid weight can be written as the integral of Fubini-Study metric, which is bounded by the Wilson loop winding number of the two flat bands of TBLG. Based on the continuum model and the superconducting transition temperature measured in TBLG experiments, we estimated the magnitude of the superfluid weight in TBLG, and show that the topological contribution may play a significant role. The same argument can be applied to other Moir\'e superlattices with flat bands and non-trivial band topology. This shows that topological bands can have $T_c$ much higher than their band widths. We also note that our lower bound is consistent with the upper bound studied in Ref. \cite{hazra2018upperbound}. For topological flat bands, the upper bound in Ref. \cite{hazra2018upperbound} is around the energy gap $E_g$ between the flat bands and other bands, while our lower bound is proportional to the order parameter $\Delta$, which is derived under the assumption that $|\Delta| \ll E_g$.

%\begin{acknowledgements}
%We acknowledge helpful discussion with Nai Phuan Ong, Ali Yazdani and Guorui Chen. ZDS and BB is supported by the Department of Energy Grant No. de-sc0016239,  Grant No. noaawd1004957, Simons Investigator Grants No. ONRN00014-14-1-0330, the Packard Foundation, the Schmidt Fund for Innovative Research. The National Science Foundation EAGER and and No. NSF-MRSEC DMR- 1420541 also supported the work. BL is supported by Princeton Center for Theoretical Science at Princeton University.
%\end{acknowledgements}

\emph{Acknowledgments.}
We acknowledge helpful discussion with Nai Phuan Ong, Ali Yazdani and Guorui Chen. ZS and BB are supported by the Department of Energy Grant No. {DE-SC}0016239, the National Science Foundation EAGER Grant No. DMR 1643312, Simons Investigator Grants No. 404513, ONR No. N00014-14-1-0330, NSF-MRSEC No. DMR-142051, the Packard Foundation, the Schmidt Fund for Innovative Research. BL is supported by Princeton Center for Theoretical Science at Princeton University.

\bibliography{reference}
%%Supplementary Material starts here.

\begin{widetext}

\beginsupplement

\section*{Supplementary Material}

The supplementary material sections are organized as follows:
\begin{itemize}

\item Sec. \ref{sec:swdef}: MEAN FIELD THEORY, BOGOLIUBOV QUASIPARTICLES, AND THE ZERO TEMPERATURE GRAND POTENTIAL

\item Sec. \ref{sec:mfsw}: SUPERFLUID WEIGHT UNDER MEAN-FIELD APPROXIMATION

\item Sec. \ref{sec:fbsw}: SUPERFLUID WEIGHT WITHIN FLAT BANDS AND RELATION TO THE QUANTUM METRIC

\item Sec. \ref{sec:metric}: QUANTUM GEOMETRIC TENSOR, FUBINI-STUDY METRIC AND BERRY CURVATURE: LOWER BOUNDS

\item Sec. \ref{sec:lowerbound}: BOUNDS ON THE QUANTUM METRIC FOR DIFFERENT SYSTEMS: SSH CHAIN, CHERN INSULATOR, AND $C_{2z}T$ FRAGILE TOPOLOGY

\item Sec. \ref{sec:sctblg}: SUPERCONDUCTIVITY IN TBLG

\item Sec. \ref{sec:ftbkt}: FINITE TEMPERATURE STUDY AND BEREZINSKII-KOSTERLITZ-THOULESS TRANSITION

\item Sec. \ref{sec:validity}: THE VALIDITY OF BCS WAVE FUNCTION IN THE FLAT BAND LIMIT

\end{itemize}
In the case without specification, we set $\hbar = e = k_B =1$ in supplementary material for convenience.

\section{Mean field theory, Bogoliubov quasiparticles and the zero temperature free energy}\label{sec:swdef}
In this section, we review the BdG mean field Hamiltonian of a superconductor in the presence of an external gauge field. Before we consider the interaction between electrons, we write down the free fermion Hamiltonian:
\begin{equation}
    H_0 = \sum_{\mathbf{k}}\psi_{\mathbf{k}\alpha}^\dagger\mathcal{H}_{\alpha\beta}(\mathbf{k})\psi_{\mathbf{k}\beta}\,,
\end{equation}
where $\psi_{\mathbf{k}}$ is a spinor with $N_B$ components labeled by $\alpha$. Here $\alpha$ can stand for the local orbitals in tight binding models, or reciprocal lattices in continuum models, such as the $\mathbf{Q}$ lattice in TBLG \cite{bistritzer2011,songz2018}. The spin index has also been included in $\alpha$. The Bloch Hamiltonian $\mathcal{H}(\mathbf{k})$ can be diagonalized by a unitary matrix $U(\mathbf{k})$ as $\varepsilon_\mathbf{k} = U^\dagger(\mathbf{k})\mathcal{H}(\mathbf{k})U(\mathbf{k})$. If a time dependent uniform gauge field $\mathbf{q} = e\mathbf{A}$ - which corresponds to a uniform electric field -, is applied, we can obtain the Hamiltonian with Peierls substitution by shifting $\mathbf{k}$ to $\mathbf{k}-\mathbf{q}$:
\begin{equation}
  H_0(\mathbf{q}) = \sum_{\mathbf{k}}\psi^\dagger_{\mathbf{k}\alpha}\mathcal{H}_{\alpha\beta}(\mathbf{k}-\mathbf{q}) \psi_{\mathbf{k}\beta}\,.
\end{equation}

The attractive interaction between the electrons leads to pairing terms  such as $\psi^\dagger\psi^\dagger$ appear in the Hamiltonian under mean-field approximation. We consider the case without FFLO order, hence only the pairing between electrons with momentum $\mathbf{k}$ and $-\mathbf{k}$ exists. In general, the pairing term takes the form:
\begin{equation}
  H_{\rm pair} = \frac12\sum_{\mathbf{k}}\left(\Delta_{\alpha\beta}(\mathbf{k})\psi^\dagger_{\mathbf{k}\alpha}\psi^\dagger_{-\mathbf{k}\beta} + {\rm h.c.}\right)\,,
\end{equation}
where $\Delta(\mathbf{k}) = -\Delta^{\rm T}(-\mathbf{k})$ because of particle hole redundancy. Since we have used Peierls substitution, the effect of gauge field has been included in kinetic term, and the pairing term will not depend on $\mathbf{q}$. The Hamiltonian can be written as a bilinear form of a Nambu spinor $\Psi_\mathbf{k} = \left(\psi_{\mathbf{k}1},\psi_{\mathbf{k}2},\cdots,\psi_{-\mathbf{k}1}^\dagger,\psi_{-\mathbf{k}2}^\dagger,\cdots\right)^{\rm T}$:
\begin{align}
  H_{\rm MF}(\mathbf{q}) =& \sum_{\mathbf{k}\alpha\beta}\left[\mathcal{H}_{\alpha\beta}(\mathbf{k}-\mathbf{q}) - \mu \delta_{\alpha\beta}\right]\psi^\dagger_{\mathbf{k}\alpha} \psi_{\mathbf{k}\beta} +\frac12 \sum_{\mathbf{k}\alpha\beta}\left(\Delta_{\alpha\beta}(\mathbf{k})\psi^\dagger_{\mathbf{k}\alpha}\psi^\dagger_{-\mathbf{k}\beta} + {\rm h.c.}\right) \nonumber\\
  =& \sum_{\mathbf{k}\alpha\beta}\left[\mathcal{H}_{\alpha\beta}(\mathbf{k}-\mathbf{q}) - \mu \delta_{\alpha\beta}\right]\frac12\left(\psi^\dagger_{\mathbf{k}\alpha} \psi_{\mathbf{k}\beta} - \psi_{\mathbf{k}\beta}\psi^\dagger_{\mathbf{k}\alpha} + \delta_{\alpha\beta}\right) + \frac12 \sum_{\mathbf{k}\alpha\beta}\left(\Delta_{\alpha\beta}(\mathbf{k})\psi^\dagger_{\mathbf{k}\alpha}\psi^\dagger_{-\mathbf{k}\beta} + {\rm h.c.}\right) \nonumber\\
  =& \frac12 \sum_{\mathbf{k}\alpha\beta} \left[\mathcal{H}_{\alpha\beta}(\mathbf{k}-\mathbf{q}) - \mu \delta_{\alpha\beta}\right]\psi^\dagger_{\mathbf{k}\alpha} \psi_{\mathbf{k}\beta} - \frac12 \sum_{\mathbf{k}\alpha\beta}\left[\mathcal{H}_{\alpha\beta}(-\mathbf{k}-\mathbf{q}) - \mu \delta_{\alpha\beta}\right]\psi_{-\mathbf{k}\beta} \psi^\dagger_{-\mathbf{k}\alpha} \nonumber\\
  &+ \frac12 \sum_{\mathbf{k}\alpha\beta}\delta_{\alpha\beta}\left[\mathcal{H}_{\alpha\beta}(\mathbf{k}-\mathbf{q}) - \mu \delta_{\alpha\beta}\right] + \frac12 \sum_{\mathbf{k}\alpha\beta}\left(\Delta_{\alpha\beta}(\mathbf{k})\psi^\dagger_{\mathbf{k}\alpha}\psi^\dagger_{-\mathbf{k}\beta} + {\rm h.c.}\right) \nonumber\\
  =&\frac12 \sum_{\mathbf{k}}\Psi_\mathbf{k}^\dagger\left(
    \begin{array}{cc}
      \mathcal{H}(\mathbf{k}-\mathbf{q}) - \mu & \Delta(\mathbf{k}) \\ \Delta^\dagger(\mathbf{k}) & -\mathcal{H}^{\rm T}(-\mathbf{k}-\mathbf{q}) + \mu
    \end{array}
  \right)\Psi_{\mathbf{k}} + \frac12 \sum_\mathbf{k}{\rm Tr\,}(\mathcal{H}(\mathbf{k}-\mathbf{q})-\mu) \,.\label{eqn:bdgderivation}
\end{align}
If the free fermion Hamiltonian can be diagonalized by matrix $U(\mathbf{k})$ as $\varepsilon_{\mathbf{k}} = U^\dagger(\mathbf{k})\mathcal{H}(\mathbf{k})U(\mathbf{k})$ where $\varepsilon_\mathbf{k}$ is a diagonal matrix, we will be able to use it to diagonalize the diagonal blocks of Eq. (\ref{eqn:bdgderivation}):
\begin{align}
  H_{\rm MF}(\mathbf{q}) =& \frac{1}{2}\sum_{\mathbf{k}}d^\dagger_{\mathbf{k}}\left(
    \begin{array}{cc}
      \varepsilon_{\mathbf{k}-\mathbf{q}} - \mu & U^\dagger(\mathbf{k}-\mathbf{q})\Delta(\mathbf{k}) U^*(-\mathbf{k}-\mathbf{q}) \\ U^{\rm T}(-\mathbf{k}-\mathbf{q})\Delta^\dagger(\mathbf{k}) U(\mathbf{k}-\mathbf{q}) & -\varepsilon_{-\mathbf{k}-\mathbf{q}} + \mu
    \end{array}
  \right)d_{\mathbf{k}}+\frac12 \sum_\mathbf{k}{\rm Tr\,}(\varepsilon_{\mathbf{k}-\mathbf{q}}-\mu)\nonumber\\
  & = \frac{1}{2}\sum_{\mathbf{k}} d^\dagger_\mathbf{k}\mathscr{H}_\mathbf{k}(\mathbf{q})d_\mathbf{k}+\frac12 \sum_\mathbf{k}{\rm Tr\,}(\varepsilon_{\mathbf{k}-\mathbf{q}}-\mu)\\
  \mathscr{H}_\mathbf{k}(\mathbf{q}) &= \left(
    \begin{array}{cc}
      \varepsilon_{\mathbf{k}-\mathbf{q}} - \mu & U^\dagger(\mathbf{k}-\mathbf{q})\Delta(\mathbf{k}) U^*(-\mathbf{k}-\mathbf{q}) \\ U^{\rm T}(-\mathbf{k}-\mathbf{q})\Delta^\dagger(\mathbf{k}) U(\mathbf{k}-\mathbf{q}) & -\varepsilon_{-\mathbf{k}-\mathbf{q}} + \mu
    \end{array}
  \right)\,,\label{eqn:bdgmatrix}\\
  d_{\mathbf{k}} &= \left( \begin{array}{cc} U^\dagger(\mathbf{k}-\mathbf{q}) & \\ & U^{\rm T}(-\mathbf{k}-\mathbf{q}) \end{array}\right)\Psi_{\mathbf{k}}\,,\nonumber
\end{align}

By diagonalizing the Bogoliubov-de-Gennes Hamiltonian $\mathscr{H}_\mathbf{k}(\mathbf{q})$, the quasiparticle spectrum of the superconductor can be obtained. Assume the matrix $\mathscr{H}_\mathbf{k}(\mathbf{q})$ is diagonalized by unitary matrix $W_\mathbf{k}(\mathbf{q})$, we will have the following equation:
\begin{equation}\label{eqn:bogoliubov}
  E_{\mathbf{k}}(\mathbf{q}) = W^\dagger_\mathbf{k}(\mathbf{q})\mathscr{H}_\mathbf{k}(\mathbf{q})W_\mathbf{k}(\mathbf{q})\,,
\end{equation}
where $E_{\mathbf{k}}(\mathbf{q})$ is a diagonal matrix and its eigenvalues $E_{\mathbf{k}n}(\mathbf{q})$ give us the quasiparticle spectrum. When $\mathbf{q} = 0$, the matrix $W(0)$ is the Bogoliubov quasiparticle coefficients:
\begin{equation}
  W_\mathbf{k}(0) = \left(
    \begin{array}{cc}
      \mathcal{U}_\mathbf{k} & \mathcal{V}_\mathbf{k} \\ \mathcal{V}^*_{-\mathbf{k}} & \mathcal{U}^*_{-\mathbf{k}}
    \end{array}
  \right)\,,
\end{equation}
and because of the unitarity of matrix $W_\mathbf{k}(0)$, the blocks satisfy $\mathcal{U}_{\mathbf{k}}\mathcal{U}^\dagger_{\mathbf{k}}+\mathcal{V}_{\mathbf{k}}\mathcal{V}^\dagger_{\mathbf{k}}= 1,~\mathcal{U}_{\mathbf{k}}\mathcal{V}^{\rm T}_{-\mathbf{k}} + \mathcal{V}_{\mathbf{k}}\mathcal{U}^{\rm T}_{-\mathbf{k}}=0$. These matrices will be helpful for us to simplify the expression of superfluid weight later. At zero temperature, the free energy is given by the summation of all the negative eigenvalues of $\mathscr{H}_\mathbf{k}(\mathbf{q})$ at every momentum $\mathbf{k}$ in the BZ. More precisely, the free energy at zero temperature can be expressed as
\begin{align}
  \Omega(\mathbf{q}) &= \frac{1}{2}\sum_{\mathbf{k}, E_{\mathbf{k}n}(\mathbf{q})<0 } E_{\mathbf{k}n}(\mathbf{q}) + \frac12 \sum_{\mathbf{k}}{\rm Tr}\,(\varepsilon_{\mathbf{k}-\mathbf{q}} -\mu)\nonumber\\
  &=  \frac14\sum_{\mathbf{k}}\left(\sum_{E_{\mathbf{k}n}(\mathbf{q})<0}E_{\mathbf{k}n}(\mathbf{q})+\sum_{E_{\mathbf{k}n}(\mathbf{q})>0}E_{\mathbf{k}n}(\mathbf{q})+\sum_{E_{\mathbf{k}n}(\mathbf{q})<0}E_{\mathbf{k}n}(\mathbf{q})-\sum_{E_{\mathbf{k}n}(\mathbf{q})>0}E_{\mathbf{k}n}(\mathbf{q})\right) + \frac12 \sum_{\mathbf{k}}{\rm Tr}\,(\varepsilon_{\mathbf{k}-\mathbf{q}} -\mu)\nonumber\\
  &= \frac14\sum_{\mathbf{k},n}E_{\mathbf{k}n}(\mathbf{q}) - \frac14\sum_{\mathbf{k},n}|E_{\mathbf{k}n}(\mathbf{q})|+\frac12 \sum_{\mathbf{k}}{\rm Tr}\,(\varepsilon_{\mathbf{k}-\mathbf{q}} -\mu)\nonumber\\
  &=\frac14\sum_\mathbf{k}{\rm Tr}\,\mathscr{H}_\mathbf{k}(\mathbf{q})-\frac14\sum_{\mathbf{k},n}|E_{\mathbf{k}n}(\mathbf{q})|+\frac12 \sum_{\mathbf{k}}{\rm Tr}\,(\varepsilon_{\mathbf{k}-\mathbf{q}} -\mu)\nonumber\\
  &= \frac14\left(\sum_\mathbf{k}{\rm Tr}\,(\varepsilon_\mathbf{k-\mathbf{q}}-\mu)-\sum_{\mathbf{k}}{\rm Tr}\,(\varepsilon_\mathbf{-k-\mathbf{q}}-\mu)\right)-\frac14\sum_{\mathbf{k},n}|E_{\mathbf{k}n}(\mathbf{q})|+\frac12 \sum_{\mathbf{k}}{\rm Tr}\,(\varepsilon_{\mathbf{k}-\mathbf{q}} -\mu)\nonumber\\
  &=-\frac14\sum_{\mathbf{k},n}|E_{\mathbf{k}n}(\mathbf{q})|+\frac12 \sum_{\mathbf{k}}{\rm Tr}\,(\varepsilon_{\mathbf{k}-\mathbf{q}} -\mu)\,,
\end{align}
where $\rm Tr(\cdots)$ represents the trace over the energy bands. Because of the particle hole redundancy, the eigenvalues of matrix $\mathscr{H}_\mathbf{k}(0)$ are related with each other by $E_{\mathbf{k} (N_B + n)}(0) = -E_{-\mathbf{k}n}(0) < 0$ when $1\leq n\leq N_B$ in which $N_B$ is the number of free electron bands. 

\section{Superfluid weight under mean-field approximation}\label{sec:mfsw}
For completeness, we reproduce the derivation of the superfluid weight. Our procedure is extremely similar to the derivation in Ref. \cite{Peotta2015}. For simplicity, the off-diagonal block of $\mathscr{H}_\mathbf{k}(\mathbf{q})$ is denoted by $\mathcal{D}_\mathbf{k}(\mathbf{q})$: 
\begin{equation}
  \mathcal{D}_\mathbf{k}(\mathbf{q}) = U^\dagger(\mathbf{k}-\mathbf{q}) \Delta(\mathbf{k}) U^* (-\mathbf{k}-\mathbf{q})\,,
\end{equation}
where $U(\mathbf{k})$ is the unitary matrix which diagonalize the original noninteracting Hamiltonian. $D_s$ is defined to be the second order expansion coefficients of free energy, so it can be expressed as the second order derivative of $\Omega(\mathbf{q})$:
\begin{equation}
  [D_s]_{ij}=\frac1V\frac{\partial^2 \Omega(\mathbf{q})}{\partial{q_i}\partial{q_j}}\Big{|}_{\mathbf{q}=0} = -\frac{1}{4V} \sum_{\mathbf{k}n} {\rm sgn}(E_{\mathbf{k}n}(\mathbf{q}) ) \partial_{q_i}\partial_{q_j}E_{\mathbf{k}n}(\mathbf{q}) ]\Big{|}_{\mathbf{q}=0} + \frac{1}{2V} \sum_{\mathbf{k}}{\rm Tr}\,\partial_{k_i}\partial_{k_j}\varepsilon_{\mathbf{k}} \,.
\end{equation}
The first order derivative of the quasiparticle energy $E_{\mathbf{k}n}(\mathbf{q})$ is given by:
\begin{align}\label{eqn:feynmanhellmanresult}
  \partial_{q_i}E_{\mathbf{k}n}(\mathbf{q}) &= \sum_{\alpha,\beta}\partial_{q_i}\left(W^*_{\mathbf{k}}(\mathbf{q})_{\alpha,n}\mathscr{H}_{\mathbf{k}}(\mathbf{q})_{\alpha,\beta}W_\mathbf{k}(\mathbf{q})_{\beta,n}\right)\nonumber\\
  & = \sum_{\alpha,\beta}W^*_{\mathbf{k}}(\mathbf{q})_{\alpha,n}\partial_{q_i}\mathscr{H}_{\mathbf{k}}(\mathbf{q})_{\alpha,\beta}W_\mathbf{k}(\mathbf{q})_{\beta,n} = \left[W^\dagger_\mathbf{k}(\mathbf{q})\partial_{q_i}\mathscr{H}_\mathbf{k}(\mathbf{q})W_\mathbf{k}(\mathbf{q})\right]_{n,n}\,,
\end{align}
here we use Feynman-Hellman theorem to derive the second equality. Further we take the second order derivative of Eq. (\ref{eqn:feynmanhellmanresult}), and it becomes:
\begin{equation}\label{eqn:sfweightterms}
  \partial_{q_i}\partial_{q_j}E_{\mathbf{k}n}(\mathbf{q}) = [W^\dagger_{\mathbf{k}}(\mathbf{q})\partial_{q_i}\partial_{q_j}\mathscr{H}_{\mathbf{k}}(\mathbf{q})W_{\mathbf{k}}(\mathbf{q})]_{n,n}+ [\partial_{q_i}W^\dagger_{\mathbf{k}}(\mathbf{q}) \partial_{q_j} \mathscr{ H }_{\mathbf{k}}(\mathbf{q})W_{\mathbf{k}}(\mathbf{q})]_{n,n} + [W^\dagger_{\mathbf{k}}(\mathbf{q})\partial_{q_j}\mathscr{H}_{\mathbf{k}}(\mathbf{q})\partial_{q_i}W_{\mathbf{k}}(\mathbf{q})]_{n,n}\,.
\end{equation}
This expression can be further simplified. Suppose $W_\mathbf{k}(\mathbf{q})_{\alpha,n}$ is the $n$-th eigenvector of matrix $\mathscr{H}_\mathbf{k}(\mathbf{q})_{\alpha,\beta}$ where $\mathbf{q}$ is a parameter, then by definition we have $\sum_\beta\mathscr{H}_\mathbf{k}(\mathbf{q})_{\alpha,\beta}W_\mathbf{k}(\mathbf{q})_{\beta,n} = E_{\mathbf{k}n}(\mathbf{q})W_\mathbf{k}(\mathbf{q})_{\alpha,n}$. Because $\mathscr{H}_\mathbf{k}(\mathbf{q})$ is a hermitian matrix, we will have the following equation when $n\neq m$:
$$
\sum_{\alpha\beta} W^*_\mathbf{k}(\mathbf{q})_{\alpha,n}\mathscr{H}_{\mathbf{k}}(\mathbf{q})_{\alpha,\beta}W_\mathbf{k}(\mathbf{q})_{\beta,m} = 0\,.
$$
Now we take derivative $\partial_{q_i}$ on both sides of this equation, and we obtain (for $n \ne m$) 
\begin{align*}
0 &= \sum_{\alpha\beta}\Big{(} W^*_\mathbf{k}(\mathbf{q})_{\alpha,n}\partial_{q_i}\mathscr{H}_{\mathbf{k}}(\mathbf{q})_{\alpha,\beta}W_\mathbf{k}(\mathbf{q})_{\beta,m} + \partial_{q_i}W^*_\mathbf{k}(\mathbf{q})_{\alpha,n}\mathscr{H}_{\mathbf{k}}(\mathbf{q})_{\alpha,\beta}W_\mathbf{k}(\mathbf{q})_{\beta,m} + W^*_\mathbf{k}(\mathbf{q})_{\alpha,n}\mathscr{H}_{\mathbf{k}}(\mathbf{q})_{\alpha,\beta}\partial_{q_i}W_\mathbf{k}(\mathbf{q})_{\beta,m} \Big{)}\\
&= \sum_{\alpha\beta}W^*_\mathbf{k}(\mathbf{q})_{\alpha,n}\partial_{q_i}\mathscr{H}_{\mathbf{k}}(\mathbf{q})_{\alpha,\beta}W_\mathbf{k}(\mathbf{q})_{\beta,m} +\sum_\alpha\Big{(}  E_{\mathbf{k}m}(\mathbf{q})\partial_{q_i}W^*_\mathbf{k}(\mathbf{q})_{\alpha,m}W_\mathbf{k}(\mathbf{q})_{\alpha,n} +E_{\mathbf{k}n}(\mathbf{q}) W^*_\mathbf{k}(\mathbf{q})_{\alpha,m}\partial_{q_i}W_\mathbf{k}(\mathbf{q})_{\alpha,n} \Big{)}\\
&=  \sum_{\alpha\beta}W^*_\mathbf{k}(\mathbf{q})_{\alpha,n}\partial_{q_i}\mathscr{H}_{\mathbf{k}}(\mathbf{q})_{\alpha,\beta}W_\mathbf{k}(\mathbf{q})_{\beta,m}  + (E_{\mathbf{k}m}(\mathbf{q})-E_{\mathbf{k}n}(\mathbf{q}))\sum_\alpha W^*_\mathbf{k}(\mathbf{q})_{\alpha,m}\partial_{q_i}W_\mathbf{k}(\mathbf{q})_{\alpha,n} \,,
\end{align*}
The unitarity of $W_\mathbf{k}(\mathbf{q})$ is used in the third equality. This result can also be written as
$$
  \sum_\alpha W^*_\mathbf{k}(\mathbf{q})_{\alpha,n}\partial_{q_i}W_\mathbf{k}(\mathbf{q})_{\alpha,m}  = \frac{\sum_{\alpha\beta}W^*_\mathbf{k}(\mathbf{q})_{\alpha,n}\partial_{q_i}\mathscr{H}_{\mathbf{k}}(\mathbf{q})_{\alpha,\beta}W_\mathbf{k}(\mathbf{q})_{\beta,n}}{E_{\mathbf{k}m}(\mathbf{q})-E_{\mathbf{k}n}(\mathbf{q})}\,,
$$
or equivalently
$$
[W^\dagger_\mathbf{k}(\mathbf{q})\partial_{q_i}W_\mathbf{k}(\mathbf{q})]_{n,m} = \frac{[W^\dagger_\mathbf{k}(\mathbf{q})\partial_{q_i}\mathscr{H}_\mathbf{k}(\mathbf{q})W_\mathbf{k}(\mathbf{q})]_{n,m}}{E_{\mathbf{k}m}(\mathbf{q})-E_{\mathbf{k}n}(\mathbf{q})}\,.
$$
This theorem (which is nothing else than the classic expression of the non-abelian Berry phase in terms of a sum over eigenstates) can be used to simplify the second and third terms in Eq. (\ref{eqn:sfweightterms}):
\begin{align}
  &[\partial_{q_i}W^\dagger_{\mathbf{k}}(\mathbf{q}) \partial_{q_j} \mathscr{ H }_{\mathbf{k}}(\mathbf{q})W_{\mathbf{k}}(\mathbf{q})]_{n,n} + [W^\dagger_{\mathbf{k}}(\mathbf{q})\partial_{q_j}\mathscr{H}_{\mathbf{k}}(\mathbf{q})\partial_{q_i}W_{\mathbf{k}}(\mathbf{q})]_{n,n}\nonumber\\
  =&\sum_{m}\left([\partial_{q_i}W^\dagger_{\mathbf{k}}(\mathbf{q})W_{\mathbf{k}}(\mathbf{q})]_{n,m}[W^\dagger_{\mathbf{k}}(\mathbf{q}) \partial_{q_j} \mathscr{ H }_{\mathbf{k}}(\mathbf{q})W_{\mathbf{k}}(\mathbf{q})]_{m,n} + [W^\dagger_{\mathbf{k}}(\mathbf{q}) \partial_{q_j} \mathscr{ H }_{\mathbf{k}}(\mathbf{q})W_{\mathbf{k}}(\mathbf{q})]_{n,m}[W^\dagger_{\mathbf{k}}(\mathbf{q})\partial_{q_i}W_{\mathbf{k}}(\mathbf{q})]_{m,n} \right)\nonumber\\
  =&\sum_{m,m\neq n} \frac{[W^\dagger_\mathbf{k}(\mathbf{q})\partial_{q_i}\mathscr{H}_\mathbf{k}(\mathbf{q})W_\mathbf{k}(\mathbf{q})]_{n,m}[W^\dagger_\mathbf{k}(\mathbf{q})\partial_{q_j}\mathscr{H}_\mathbf{k}(\mathbf{q})W_\mathbf{k}(\mathbf{q})]_{m,n}}{E_{\mathbf{k}n}(\mathbf{q})-E_{\mathbf{k}m}(\mathbf{q})} + (i\leftrightarrow j)\,,
  \label{eqn:3rdtermelements}
\end{align}
and now Eq. (\ref{eqn:sfweightterms}) becomes: 
\begin{equation}\label{eqn:swterms}
  \partial_{q_i}\partial_{q_j}E_{\mathbf{k}n}(\mathbf{q}) = [W^\dagger_{\mathbf{k}}(\mathbf{q})\partial_{q_i}\partial_{q_j}\mathscr{H}_{\mathbf{k}}(\mathbf{q})W_{\mathbf{k}}(\mathbf{q})]_{n,n} + \sum_{m,m\neq n} \frac{[W^\dagger_\mathbf{k}(\mathbf{q})\partial_{q_i}\mathscr{H}_\mathbf{k}(\mathbf{q})W_\mathbf{k}(\mathbf{q})]_{n,m}[W^\dagger_\mathbf{k}(\mathbf{q})\partial_{q_j}\mathscr{H}_\mathbf{k}(\mathbf{q})W_\mathbf{k}(\mathbf{q})]_{m,n}}{E_{\mathbf{k}n}(\mathbf{q})-E_{\mathbf{k}m}(\mathbf{q})} + (i\leftrightarrow j)\,,
\end{equation}
We have hence removed the derivatives of $W_\mathbf{k}(\mathbf{q})$, which in principle are not immediately obtainable. Only the derivatives of $\mathscr{H}_\mathbf{k}(\mathbf{q})$ are left in the expressions, and these are immediately obtained. The first order derivative of $\mathscr{H}_\mathbf{k}(\mathbf{q})$ is given by:
\begin{equation}\label{eqn:1stderivative}
  \partial_{q_i}\mathscr{H}_{\mathbf{k}}(\mathbf{q})\Big{|}_{\mathbf{q}=0} = \left(
    \begin{array}{cc}
      -\partial_{k_i} \varepsilon_{\mathbf{k}} & \partial_{q_i}\mathcal{D}_{\mathbf{k}}(\mathbf{q})\Big{|}_{\mathbf{q}=0}\\
      \partial_{q_i}\mathcal{D}^\dagger_{\mathbf{k}}(\mathbf{q})\Big{|}_{\mathbf{q}=0} & -\partial_{k_i}\varepsilon_{-\mathbf{k}}
    \end{array}
  \right)\,.
\end{equation}
Similarly the second order derivative of matrix $\mathscr{H}_\mathbf{k}(\mathbf{q})$ will be:
\begin{equation}\label{eqn:2ndderivative}
  \partial_{q_i}\partial_{q_j} \mathscr{H}_{\mathbf{k}}(\mathbf{q})\Big{|}_{\mathbf{q}=0} = \left(
    \begin{array}{cc}
      \partial_{k_i}\partial_{k_j}\varepsilon_{\mathbf{k}} & \partial_{q_i}\partial_{q_j}\mathcal{D}_{\mathbf{k}}(\mathbf{q})\Big{|}_{\mathbf{q}=0} \\
      \partial_{q_i}\partial_{q_j}\mathcal{D}^\dagger_{\mathbf{k}}(\mathbf{q})\Big{|}_{\mathbf{q}=0} & -\partial_{k_i}\partial_{k_j}\varepsilon_{-\mathbf{k}}
    \end{array}
  \right)
\end{equation}
We substitute Eqs. (\ref{eqn:1stderivative}) and (\ref{eqn:2ndderivative}) into Eq. (\ref{eqn:swterms}) to get the result of superfluid weight. Now only the Bogoliubov coefficients $W_\mathbf{k}(0)$ are necessary. By using Eq. (\ref{eqn:1stderivative}) and the first term of Eq. (\ref{eqn:swterms}), we obtain the following two terms of superfluid weight:
\begin{align}
  [D_s^{(1)}]_{ij} &= \frac{1}{2V}\sum_{\mathbf{k}} {\rm Tr}\left(\mathcal{V}_{\mathbf{k}} \mathcal{V}^\dagger_{\mathbf{k}}\partial_{k_i}\partial_{k_j}\varepsilon_{\mathbf{k}}+ \mathcal{V}^*_{-\mathbf{k}} \mathcal{V}^{\rm T}_{-\mathbf{k}} \partial_{k_i}\partial_{k_j}\varepsilon_{-\mathbf{k}} \right)\label{eqn:swterm1} \\
  [D_s^{(2)}]_{ij} &= \frac{1}{4V}\sum_{\mathbf{k}} {\rm Tr}\left[\left(\mathcal{U}_{\mathbf{k}}\mathcal{V}^{\rm T}_{-\mathbf{{k}}}-\mathcal{V}_{\mathbf{k}}\mathcal{U}^{\rm T}_{-\mathbf{k}}\right)\partial_{q_i}\partial_{q_j}\mathcal{D}_{\mathbf{k}}^\dagger(\mathbf{q})\Big{|}_{\mathbf{q}=0} + \left(\mathcal{U}^{*}_{-\mathbf{{k}}}\mathcal{V}^\dagger_{\mathbf{k}}-\mathcal{V}^*_{-\mathbf{k}}\mathcal{U}^\dagger_{\mathbf{k}}\right)\partial_{q_i}\partial_{q_j}\mathcal{D}_{\mathbf{k}}(\mathbf{q})\Big{|}_{\mathbf{q}=0}\right]\label{eqn:swterm2}\,.
\end{align}
The contribution from $\frac12 \sum_\mathbf{k}\partial_{k_i}\partial_{k_j}\varepsilon_{\mathbf{k}}$ has been included in $[D_s]^{(1)}$. Similarly, by substituting Eq. (\ref{eqn:3rdtermelements}) into Eq. (\ref{eqn:swterms}), we get the third term of the superfluid weight:
\begin{equation}\label{eqn:swterm3ori}
  [D_s^{(3)}]_{ij} = -\frac{1}{4V}\sum_{\mathbf{k}}\sum_{n=1}^{2N_B}\sum_{m=1,m\neq n}^{2N_B}{\rm sgn}(N_B-n + 1/2)\left\{ \frac{[W^\dagger_{\mathbf{k}}\partial_{q_i}\mathscr{H}_{\mathbf{k}}(\mathbf{q})W_{\mathbf{k}}]_{n,m}[W^\dagger_{\mathbf{k}}\partial_{q_j}\mathscr{H}_{\mathbf{k}}(\mathbf{q})W_{\mathbf{k}}]_{m,n}}{E_{\mathbf{k}n}(0)-E_{\mathbf{k}m}(0)} + (i\leftrightarrow j)\right\}\,.
\end{equation}
We use Eqs. (\ref{eqn:bogoliubov}) and (\ref{eqn:1stderivative}) to express $W^\dagger_{\mathbf{k}}\partial_{q_i}\mathscr{H}_{\mathbf{k}}(\mathbf{q})W_{\mathbf{k}}|_{\mathbf{q}=0}$ by the following block matrix:
\begin{align}
  W^\dagger_{\mathbf{k}}\partial_{q_i}\mathscr{H}_{\mathbf{k}}(\mathbf{q})W_{\mathbf{k}}\Big{|}_{\mathbf{q}=0}& = \left(
  \begin{array}{cc}
    P_i(\mathbf{k}) & Q_i(\mathbf{k})\\
    Q^\dagger_i(\mathbf{k}) & R_i(\mathbf{k})
  \end{array}
  \right)\\
  P_i(\mathbf{k}) &= -\mathcal{U}_{\mathbf{k}}^\dagger\partial_{k_i}\varepsilon_{\mathbf{k}}\mathcal{U}_{\mathbf{k}} +\mathcal{V}_{-\mathbf{k}}^{\rm T}\partial_{q_i}\mathcal{D}_{\mathbf{k}}^\dagger(\mathbf{q})\Big{|}_{\mathbf{q}=0}\mathcal{U}_{\mathbf{k}} + \mathcal{U}^\dagger_{\mathbf{k}}\partial_{q_i}\mathcal{D}_{\mathbf{k}}(\mathbf{q})\Big{|}_{\mathbf{q}=0} \mathcal{V}^*_{-\mathbf{k}} - \mathcal{V}_{-\mathbf{k}}^{\rm T}\partial_{k_i}\varepsilon_{-\mathbf{k}}\mathcal{V}^*_{-\mathbf{k}}\\
  Q_i(\mathbf{k}) &= -\mathcal{U}^\dagger_{\mathbf{k}}\partial_{k_i}\varepsilon_{\mathbf{k}}\mathcal{V}_{\mathbf{k}} +\mathcal{V}^{\rm T}_{-\mathbf{k}}\partial_{q_i}\mathcal{D}_{\mathbf{k}}^\dagger(\mathbf{q})\Big{|}_{\mathbf{q}=0}\mathcal{V}_{\mathbf{k}} +\mathcal{U}^\dagger_{\mathbf{k}}\partial_{q_i}\mathcal{D}_{\mathbf{k}}(\mathbf{q})\Big{|}_{\mathbf{q}=0}\mathcal{U}^*_{-\mathbf{k}} -\mathcal{V}^{\rm T}_{-\mathbf{k}}\partial_{k_i}\varepsilon_{-\mathbf{k}}\mathcal{U}^*_{-\mathbf{k}} \\
  R_i(\mathbf{k}) &= -\mathcal{V}^\dagger_{\mathbf{k}}\partial_{k_i}\varepsilon_{\mathbf{k}}\mathcal{V}_{\mathbf{k}} + \mathcal{U}^{\rm T}_{-\mathbf{k}}\partial_{q_i}\mathcal{D}_{\mathbf{k}}^\dagger(\mathbf{q})\Big{|}_{\mathbf{q}=0} \mathcal{V}_{\mathbf{k}} +\mathcal{V}^\dagger_{\mathbf{k}} \partial_{q_i}\mathcal{D}_{\mathbf{k}}(\mathbf{q})\Big{|}_{\mathbf{q}=0} \mathcal{U}^*_{-\mathbf{k}} - \mathcal{U}^{\rm T}_{-\mathbf{k}} \partial_{k_i}\varepsilon_{-\mathbf{k}} \mathcal{U}_{-\mathbf{k}}^*
\end{align}
We use these blocks to express the third term of superfluid weight (Eq. (\ref{eqn:swterm3ori})), and it can be written as
\begin{align}
  [D_s]^{(3)}_{ij} =& -\frac{1}{4V} \sum_{\mathbf{k}\in {\rm BZ}} \left\{ \sum_{n=1}^{N_B}\sum_{n'=1,n'\neq n}^{N_B}\left[\frac{[P_i(\mathbf{k})]_{n,n'}[P_j(\mathbf{k})]_{n',n}}{E_{\mathbf{k}n}(0)-E_{\mathbf{k}n'}(0)} + \frac{[P_j(\mathbf{k})]_{n,n'}[P_i(\mathbf{k})]_{n',n}}{E_{\mathbf{k}n}(0)-E_{\mathbf{k}n'}(0)}-\frac{[R_i(\mathbf{k})]_{n,n'}[R_j(\mathbf{k})]_{n',n}}{E_{\mathbf{k}(n+N_B)}(0)-E_{\mathbf{k}(n'+N_B)}(0)}\right. \right.\nonumber\\
  &\left. - \frac{[R_j(\mathbf{k})]_{n,n'}[R_i(\mathbf{k})]_{n',n}}{E_{\mathbf{k}(n+N_B)}(0)-E_{\mathbf{k}(n'+N_B)}(0)} \right] + \sum_{n,n'=1}^{N_B} \left[\frac{[Q_i(\mathbf{k})]_{n,n'}[Q^\dagger_j(\mathbf{k})]_{n',n}}{E_{\mathbf{k}n}(0)-E_{\mathbf{k}(n'+N_B)}(0)} + \frac{[Q_j(\mathbf{k})]_{n,n'}[Q_i^\dagger(\mathbf{k})]_{n',n}}{E_{\mathbf{k}n}(0)-E_{\mathbf{k}(n'+N_B)}(0)} \right.\nonumber\\
  &\left. \left.- \frac{[Q^\dagger_i(\mathbf{k})]_{n,n'}[Q_j(\mathbf{k})]_{n',n}}{E_{\mathbf{k}(n+N_B)}(0)-E_{\mathbf{k}n'}(0)}-\frac{[Q^\dagger_j(\mathbf{k})]_{n,n'}[Q_i(\mathbf{k})]_{n',n}}{E_{\mathbf{k}(n+N_B)}(0)-E_{\mathbf{k}n'}(0)} \right]\right\} \,.\label{eqn:3rdtermmid}
\end{align}
First we study the contribution from these $P_i(\mathbf{k})$ matrices. The first term in Eq. (\ref{eqn:3rdtermmid}) is given by:
\begin{equation}
\sum_{n=1}^{N_B}\sum_{n'=1,n'\neq n}^{N_B}\frac{[P_i(\mathbf{k})]_{n,n'}[P_j(\mathbf{k})]_{n',n}}{E_{\mathbf{k}n}(0)-E_{\mathbf{k}n'}(0)}\,,
\end{equation}
and we notice it has two dummy indices $n$ and $n'$. If we simply switch them ($n\leftrightarrow n'$), it will become
\begin{equation}
\sum_{n=1}^{N_B}\sum_{n'=1,n'\neq n}^{N_B}\frac{[P_i(\mathbf{k})]_{n',n}[P_j(\mathbf{k})]_{n,n'}}{E_{\mathbf{k}n'}(0)-E_{\mathbf{k}n}(0)} = - \sum_{n=1}^{N_B}\sum_{n'=1,n'\neq n}^{N_B} \frac{[P_j(\mathbf{k})]_{n,n'}[P_i(\mathbf{k})]_{n',n}}{E_{\mathbf{k}n}(0)-E_{\mathbf{k}n'}(0)}\,,
\end{equation}
and it differs from the second term in Eq. (\ref{eqn:3rdtermmid}) only by a minus sign. Therefore the summation of the first and second terms will be zero, and all $P_i(\mathbf{k})$ matrices will not appear in the final expression. Similarly, we can prove $R_i(\mathbf{k})$ will also disappear. Only matrices $Q_i(\mathbf{k})$ will be left. By using the particle hole symmetry $E_{\mathbf{k}n}(0) = - E_{-\mathbf{k}(n+N_B)}$, $[D_s]^{(3)}_{ij}$ can be expressed as:
\begin{align}
 [D_s]^{(3)}_{ij} =& \sum_{\mathbf{k}\in{\rm BZ}}\sum_{n,n'=1}^{N_B} \left[\frac{[Q_i(\mathbf{k})]_{n,n'}[Q^\dagger_j(\mathbf{k})]_{n',n}}{E_{\mathbf{k}n}(0)-E_{\mathbf{k}(n'+N_B)}(0)} + \frac{[Q_j(\mathbf{k})]_{n,n'}[Q_i^\dagger(\mathbf{k})]_{n',n}}{E_{\mathbf{k}n}(0)-E_{\mathbf{k}(n'+N_B)}(0)} - \frac{[Q^\dagger_i(\mathbf{k})]_{n,n'}[Q_j(\mathbf{k})]_{n',n}}{E_{\mathbf{k}(n+N_B)}(0)-E_{\mathbf{k}n'}(0)}-\frac{[Q^\dagger_j(\mathbf{k})]_{n,n'}[Q_i(\mathbf{k})]_{n',n}}{E_{\mathbf{k}(n+N_B)}(0)-E_{\mathbf{k}n'}(0)} \right]\nonumber\\
 &= -\frac{1}{2V}\sum_{\mathbf{k}\in {\rm BZ}} \sum_{n,n'=1}^{N_B}\left[\frac{[Q_i(\mathbf{k})]_{n,n'}[Q^\dagger_j(\mathbf{k})]_{n',n}}{E_{\mathbf{k}n}(0)+E_{-\mathbf{k}n'}(0)}+\frac{[Q_j(\mathbf{k})]_{n,n'}[Q^\dagger_i(\mathbf{k})]_{n',n}}{E_{\mathbf{k}n}(0)+E_{-\mathbf{k}n'}(0)}\right]\,.\label{eqn:swterm3}
\end{align}
By adding Eqs. (\ref{eqn:swterm1}), (\ref{eqn:swterm2}) and (\ref{eqn:swterm3}) together, we finally get the general expression of superfluid weight. 
This result is accurate under mean-field approximation without FFLO order. In the following section we will see that the superfluid weight can be related to the quantum geometric tensor under the flat band approximation. Our result here is very similar to the result shown in Ref. \cite{Peotta2015}, but it is more general. We used the Nambu spinor with particle-hole redundancy, and our result can also be applied to Hamiltonians with Rashba coupling.

\section{Superfluid weight within flat bands and relation to the quantum metric}\label{sec:fbsw}
In this section we study the superfluid weight in the flat band limit. The following discussion is based on these assumptions: 1) the free fermion Hamiltonian $\mathcal{H}(\mathbf{k})$ has $N_F$ flat bands, with energy $\varepsilon_0$, and it can be diagonalized by unitary matrix $U(\mathbf{k})$; 2) the band gap between flat bands and other bands is larger than the band width of flat bands (when the bands are exactly flat, this must be true) and the pairing order parameter. Because of the large band gap, the derivation of free energy and superfluid weight can be simplified dramatically by projecting into the flat bands.

\begin{figure}[!htbp]
\centering
\includegraphics[width=8cm]{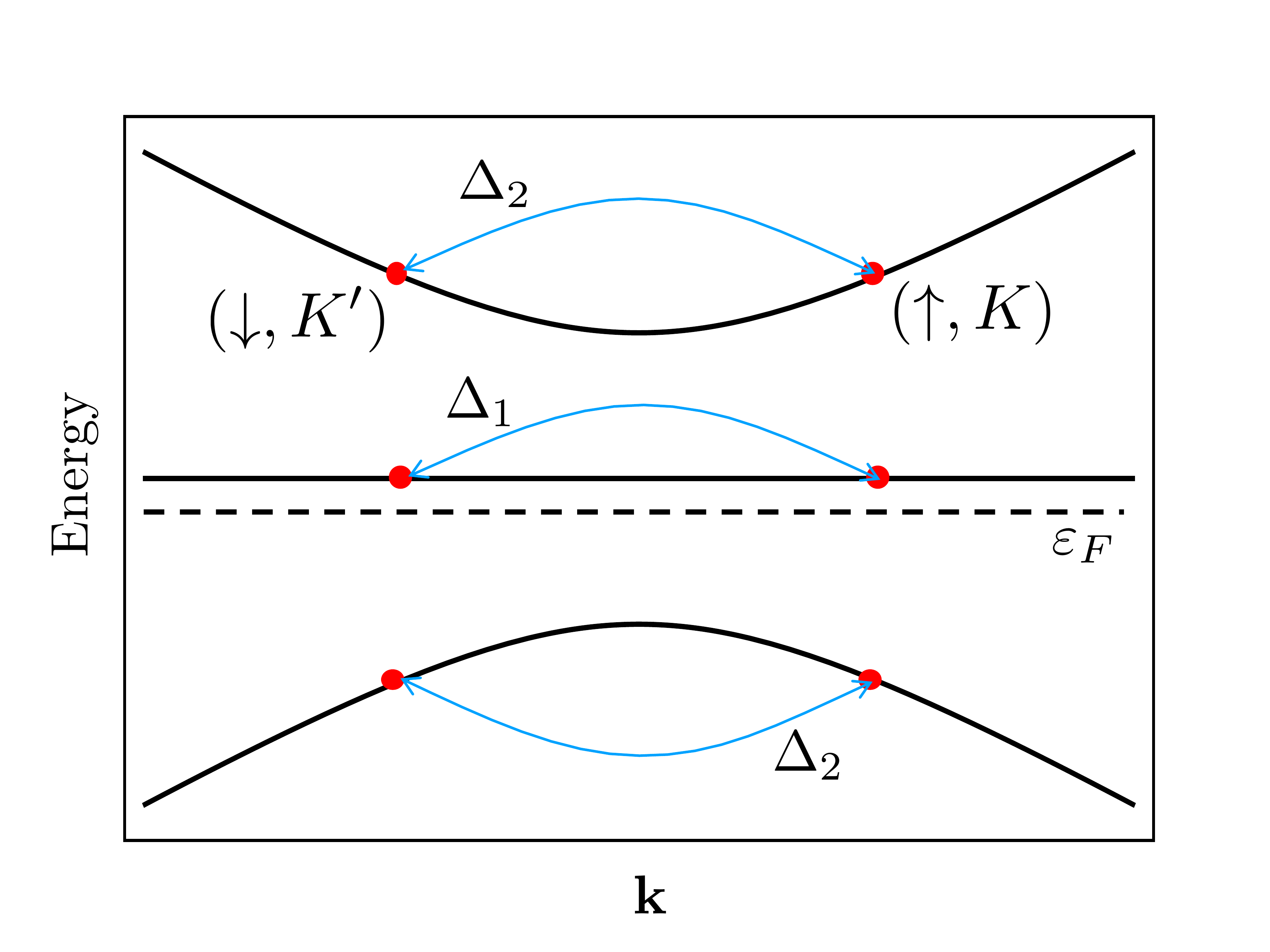}
\caption{A schematic diagram of the flat band assumption. We assume that there are $N_F$ flat bands near the Fermi level, and the gap between flat bands and other bands is large, compared to the flat band width and pairing order parameters. The pairing strength between Kramers pairs in flat bands and in other bands are $\Delta_1$ and $\Delta_2$, respectively. In the case of TBLG, the pairing happens between electrons with opposite momentum, spin, valley and in the same energy band.}
\label{fig:flatband}
\end{figure}

Now we discuss the form of pairing order parameter. It is reasonable to first assume that $\Delta(\mathbf{k})$ obeys spinful time reversal symmetry $\mathcal{T}$, as assumed in Ref. \cite{Lian2018}. This may seem irrelevant with TBLG since BM model does not have time reversal. However if we take spin $\uparrow,\downarrow$ and valley $K,K'$ into consideration, the whole system will have time reversal symmetry. The time reversal transformation can always be represented by $U_T\mathcal{K}$ in which $\mathcal{K}$ is the complex conjugation, and $U_T$ is an real antisymmetric unitary matrix acting on the orbital and spin indices. The fermion operators transform under $\mathcal{T}$ as $\mathcal{T}\psi_{\mathbf{k}i}\mathcal{T}^{-1} = (U_T)_{ij}\psi_{-\mathbf{k}j}$. If the BdG Hamiltonian has spinful time reversal symmetry, the pairing order parameter will satisfy $U_T\Delta^\dagger(\mathbf{k})U_T = \Delta(\mathbf{k})$. For simplicity we assume the order parameter has the following form:
\begin{equation}\label{eqn:deltaansatz}
  \Delta(\mathbf{k}) = \left[\Delta_1 \tilde{U}_\mathbf{k}\tilde{U}_\mathbf{k}^\dagger + \Delta_2\left(\mathds{1}-\tilde{U}_\mathbf{k}\tilde{U}_\mathbf{k}^\dagger\right)\right] U_T\,,
\end{equation}
where $\tilde{U}_\mathbf{k} = (u_1(\mathbf{k}),u_2(\mathbf{k}),\cdots,u_{N_F}(\mathbf{k}))$ stands for the projection of $U(\mathbf{k})$ into flat bands, thus $\tilde{U}_\mathbf{k}\tilde{U}_\mathbf{k}^\dagger$ is projection operator of flat bands, $\left(\mathds{1}-\tilde{U}_\mathbf{k}\tilde{U}_\mathbf{k}^\dagger\right)$ is projector of all the other bands, and $\tilde{U}^\dagger_{\mathbf{k}}\tilde{U}_\mathbf{k} = \mathds{1}_{N_F \times N_F}$. Cooper pairing happens between the Kramers pairs, and the pairing strength in flat bands is $\Delta_1$, and in other bands it is $\Delta_2$. If the BdG Hamiltonian is assumed to be time reversal invariant, then as we mentioned in last paragraph, equation $U_T\Delta^\dagger(\mathbf{k})U_T = \Delta(\mathbf{k})$ must be satisfied. This equation requires $\Delta_{1,2}$ to be real. To show this, we have the following equation
\begin{align}
  U_T\Delta^\dagger(\mathbf{k})U_T &= U_T U_T^\dagger \left[\Delta_1^* \tilde{U}_\mathbf{k}\tilde{U}_\mathbf{k}^\dagger + \Delta_2^* \left(\mathds{1}-\tilde{U}_\mathbf{k}\tilde{U}_\mathbf{k}^\dagger\right)\right] U_T\nonumber\\
  &= \left[\Delta_1^* \tilde{U}_\mathbf{k}\tilde{U}_\mathbf{k}^\dagger + \Delta_2^* \left(\mathds{1}-\tilde{U}_\mathbf{k}\tilde{U}_\mathbf{k}^\dagger\right)\right]U_T\,,
\end{align}
thus when $\Delta_{1,2}\in\mathbb{R}$, the order parameter $\Delta(\mathbf{k})$ satisfies time reversal symmetry.

As we mentioned earlier, because the band gap between flat bands and other bands is larger than other energy scales in flat bands, it is not necessary to use the tedious method in Sec. \ref{sec:mfsw} to obtain the free energy and superfluid weight. Therefore, the derivation becomes much simpler if we project the Hamiltonian into these flat bands. The BdG Hamiltonian in Eq. (\ref{eqn:bdgmatrix}) includes all the energy bands. If we only keep the columns and rows which corresponds to the flat bands, it will become:
\begin{align}
  \tilde{\mathscr{H}}_\mathbf{k}(\mathbf{q}) &= \left(
    \begin{array}{cc}
      \varepsilon_0 - \mu & \tilde{\mathcal{D}}_\mathbf{k}(\mathbf{q})\\
      \tilde{\mathcal{D}}^\dagger_\mathbf{k}(\mathbf{q}) & -\varepsilon_0 + \mu
    \end{array}
  \right)\,,\\
  \tilde{\mathcal{D}}_\mathbf{k}(\mathbf{q})&=\tilde{U}^\dagger_{\mathbf{k}-\mathbf{q}}\Delta(\mathbf{k})\tilde{U}^*_{-\mathbf{k}-\mathbf{q}} = \tilde{U}^\dagger_{\mathbf{k}-\mathbf{q}} \left[\Delta_1 \tilde{U}_\mathbf{k}\tilde{U}_\mathbf{k}^\dagger + \Delta_2\left(\mathds{1}-\tilde{U}_\mathbf{k}\tilde{U}_\mathbf{k}^\dagger\right)\right] \tilde{U}_{\mathbf{k}+\mathbf{q}}w(\mathbf{k}+\mathbf{q}) \,,
\end{align}
where $\varepsilon_0$ is the flat band energy, and $w(\mathbf{k}) = \tilde{U}^\dagger(\mathbf{k})U_T\tilde{U}^*(-\mathbf{k})$ is the time reversal sewing matrix of all the flat bands. In order to evaluate the free energy, we have to know the eigenvalues of $\tilde{\mathscr{H}}_\mathbf{k}(\mathbf{q})$. Thus we calculate the square of the projected BdG Hamiltonian:
\begin{equation}
  \tilde{\mathscr{H}}_\mathbf{k}^2(\mathbf{q}) = \left(
    \begin{array}{cc}
      (\varepsilon_0-\mu)^2 + \tilde{\mathcal{D}}_\mathbf{k}(\mathbf{q})\tilde{\mathcal{D}}^\dagger_\mathbf{k}(\mathbf{q}) & 0 \\ 0 &(\varepsilon_0-\mu)^2 + \tilde{\mathcal{D}}_\mathbf{k}^\dagger(\mathbf{q})\tilde{\mathcal{D}}_\mathbf{k}(\mathbf{q})
    \end{array}
  \right)\,.
\end{equation}
Note that it is block diagonal. We use $\lambda_{\mathbf{k}n}(\mathbf{q})$ and $\varphi_{\mathbf{k}n}(\mathbf{q})$ to denote the eigenvalues of matrices $\tilde{\mathcal{D}}_\mathbf{k}(\mathbf{q})\tilde{\mathcal{D}}^\dagger_\mathbf{k}(\mathbf{q})$ and $\tilde{\mathcal{D}}^\dagger_\mathbf{k}(\mathbf{q})\tilde{\mathcal{D}}_\mathbf{k}(\mathbf{q})$, respectively. Obviously, both $\tilde{\mathcal{D}}_\mathbf{k}(\mathbf{q})\tilde{\mathcal{D}}^\dagger_\mathbf{k}(\mathbf{q})$ and $\tilde{\mathcal{D}}^\dagger_\mathbf{k}(\mathbf{q})\tilde{\mathcal{D}}_\mathbf{k}(\mathbf{q})$ are Hermitian and semi-positive definite, we find that $\lambda_{\mathbf{k}n}(\mathbf{q})\geq 0$ and $\varphi_{\mathbf{k}n}(\mathbf{q})\geq 0$. Therefore, the eigenvalues of the projected BdG Hamiltonian satisfy $E^2_{\mathbf{k}n}(\mathbf{q}) = (\varepsilon_0 - \mu)^2 + \lambda_{\mathbf{k}n}(\mathbf{q})$ when $1 \leq n \leq N_F$ and $E^2_{\mathbf{k}n}(\mathbf{q}) = (\varepsilon_0 - \mu)^2 + \varphi_{\mathbf{k}n}(\mathbf{q})$ when $N_F + 1\leq n\leq 2N_F$. The free energy can be written can be written as the following form:
\begin{equation}
  \Omega(\mathbf{q}) =-\frac14\sum_{\mathbf{k},n}|E_{\mathbf{k}n}(\mathbf{q})| = -\frac14\sum_{\mathbf{k},n} \left( \sqrt{(\varepsilon_0-\mu)^2 + \lambda_{\mathbf{k}n}(\mathbf{q})} + \sqrt{(\varepsilon_0-\mu)^2 + \varphi_{\mathbf{k}n}(\mathbf{q})}\right)\,.
\end{equation}
Then we take the second order derivative of $\Omega(\mathbf{q})$ to get the superfluid weight:
\begin{align}
  [D_s]_{ij} =& -\frac{1}{V}\frac{\partial^2\Omega(\mathbf{q})}{\partial q_i \partial q_j}\Big{|}_{\mathbf{q}=0} \nonumber\\
  =& -\frac{1}{8V}\sum_{\mathbf{k},n}\left(\frac{\partial_{q_i}\partial_{q_j}\lambda_{\mathbf{k}n}(\mathbf{q})}{\sqrt{(\varepsilon_0-\mu)^2 + \lambda_{\mathbf{k}n}(0)}} - \frac{\partial_{q_i}\lambda_{\mathbf{k}n}(\mathbf{q}) \partial_{q_j}\lambda_{\mathbf{k}n}(\mathbf{q})}{2[(\varepsilon_0-\mu)^2+ \lambda_{\mathbf{k}n}(0)]^{3/2}} \right. \nonumber\\
  & \left.+\frac{\partial_{q_i}\partial_{q_j}\varphi_{\mathbf{k}n}(\mathbf{q})}{\sqrt{(\varepsilon_0-\mu)^2 + \varphi_{\mathbf{k}n}(0)}} - \frac{\partial_{q_i}\varphi_{\mathbf{k}n}(\mathbf{q}) \partial_{q_j}\varphi_{\mathbf{k}n}(\mathbf{q})}{2[(\varepsilon_0-\mu)^2 + \varphi_{\mathbf{k}n}(0)]^{3/2}}\right)\Bigg{|}_{\mathbf{q}=0}\,.\label{eqn:flatbandsw}
\end{align}
When $\mathbf{q} = 0$, the matrices $\tilde{\mathcal{D}}_\mathbf{k}(0)$, $\tilde{\mathcal{D}}_\mathbf{k}(0)\tilde{\mathcal{D}}_\mathbf{k}(0)^\dagger$ and $\tilde{\mathcal{D}}_\mathbf{k}(0)^\dagger\tilde{\mathcal{D}}_\mathbf{k}(0)$ will become
\begin{align}
  \tilde{\mathcal{D}}_\mathbf{k}(0) &= \tilde{U}_\mathbf{k}^\dagger \left[\Delta_1 \tilde{U}_\mathbf{k}\tilde{U}_\mathbf{k}^\dagger + \Delta_2\left(\mathds{1}-\tilde{U}_\mathbf{k}\tilde{U}_\mathbf{k}^\dagger\right)\right] \tilde{U}_{\mathbf{k}}w(\mathbf{k})\nonumber\\
  &= \left[\Delta_1 \left(\tilde{U}_\mathbf{k}^\dagger\tilde{U}_\mathbf{k}\right)^2 + \Delta_2 \left(\tilde{U}_\mathbf{k}^\dagger \tilde{U}_\mathbf{k} - \left(\tilde{U}_\mathbf{k}^\dagger \tilde{U}_\mathbf{k}\right)^2\right) \right]w(\mathbf{k})\nonumber\\
  & = \left[\Delta_1 \mathds{1}_{N_F \times N_F} + \Delta_2 \left(\mathds{1}_{N_F \times N_F} - \mathds{1}_{N_F \times N_F}\right) \right]w(\mathbf{k}) \nonumber\\
  & = \Delta_1 w(\mathbf{k})\,,
\end{align}
\begin{align}
  \tilde{\mathcal{D}}_\mathbf{k}(0)\tilde{\mathcal{D}}_\mathbf{k}(0)^\dagger& = \Delta_1 w(\mathbf{k})w^\dagger(\mathbf{k}) \Delta_1 = \Delta_1^2\,, \\
  \tilde{\mathcal{D}}^\dagger_\mathbf{k}(0)\tilde{\mathcal{D}}_\mathbf{k}(0) &= \Delta^2_1 w^\dagger(\mathbf{k})w(\mathbf{k}) = \Delta_1^2 \,.
\end{align}
Thus we must have $\lambda_{\mathbf{k}n}(0) = \varphi_{\mathbf{k}n}(0) = \Delta_1^2$, and the denominator in Eq. (\ref{eqn:flatbandsw}) will hence become a constant $E_B = \sqrt{(\varepsilon_0-\mu)^2 + \Delta_1^2}$, and the superfluid weight can be written as
\begin{equation}
  [D_s]_{ij} = -\frac{1}{8}\int \frac{d^2k}{(2\pi)^2}\sum_n \left(\frac{\partial_{q_i}\partial_{q_j}\lambda_{\mathbf{k}n}(\mathbf{q})}{E_B}+\frac{\partial_{q_i}\partial_{q_j}\varphi_{\mathbf{k}n}(\mathbf{q})}{E_B}- \frac{\partial_{q_i}\lambda_{\mathbf{k}n}(\mathbf{q}) \partial_{q_j}\lambda_{\mathbf{k}n}(\mathbf{q})}{2E_B^3}-\frac{\partial_{q_i}\varphi_{\mathbf{k}n}(\mathbf{q}) \partial_{q_j}\varphi_{\mathbf{k}n}(\mathbf{q})}{2E_B^3} \right)\Bigg{|}_{\mathbf{q}=0}\,.
\end{equation}
The first order derivative of matrices $\tilde{\mathcal{D}}_\mathbf{k}(\mathbf{q})\tilde{\mathcal{D}}^\dagger_\mathbf{k}(\mathbf{q})$ and $\tilde{\mathcal{D}}^\dagger_\mathbf{k}(\mathbf{q})\tilde{\mathcal{D}}_\mathbf{k}(\mathbf{q})$ are all equal to zero. In order to show this, we can calculate the first order derivative of $\mathcal{D}_\mathbf{k}(\mathbf{q})$ and $\mathcal{D}^\dagger_\mathbf{k}(\mathbf{q})$:
\begin{align}
  \partial_{\mathbf{q}}\left(\tilde{\mathcal{D}}_\mathbf{k}(\mathbf{q})\right)\Big{|}_{\mathbf{q} = 0} &= -\partial_\mathbf{k}\tilde{U}^\dagger_\mathbf{k} \left[\Delta_1 \tilde{U}_\mathbf{k}\tilde{U}_\mathbf{k}^\dagger + \Delta_2\left(\mathds{1}-\tilde{U}_\mathbf{k}\tilde{U}_\mathbf{k}^\dagger\right)\right]\tilde{U}_\mathbf{k}w(\mathbf{k}) + \tilde{U}^\dagger_\mathbf{k} \left[\Delta_1 \tilde{U}_\mathbf{k}\tilde{U}_\mathbf{k}^\dagger + \Delta_2\left(\mathds{1}-\tilde{U}_\mathbf{k}\tilde{U}_\mathbf{k}^\dagger\right)\right]\partial_\mathbf{k}\tilde{U}_\mathbf{k}w(\mathbf{k}) + \mathcal{D}_\mathbf{k}(0)\partial_\mathbf{k}w(\mathbf{k})\nonumber\\
  &= \Delta_1 \left( -\partial_\mathbf{k}\tilde{U}^\dagger_\mathbf{k}\tilde{U}_{\mathbf{k}}w(\mathbf{k}) + \tilde{U}^\dagger \partial_{\mathbf{k}}\tilde{U}_\mathbf{k}w(\mathbf{k}) + \partial_\mathbf{k}w(\mathbf{k})\right)\,\\
  \partial_{\mathbf{q}}\left(\tilde{\mathcal{D}}^\dagger_\mathbf{k}(\mathbf{q})\right)\Big{|}_{\mathbf{q} = 0} &=  \Delta_1 \left( -w^\dagger(\mathbf{k})\tilde{U}^\dagger_\mathbf{k}\partial_\mathbf{k}\tilde{U}_{\mathbf{k}} + w^\dagger(\mathbf{k})\partial_{\mathbf{k}}\tilde{U}^\dagger_\mathbf{k} \tilde{U}_\mathbf{k} + \partial_\mathbf{k}w^\dagger(\mathbf{k})\right)\,.
\end{align}
The first order derivative of the product $\tilde{\mathcal{D}}_\mathbf{k}(\mathbf{q})\tilde{\mathcal{D}}^\dagger_\mathbf{k}(\mathbf{q})$ and $\tilde{\mathcal{D}}^\dagger_\mathbf{k}(\mathbf{q})\tilde{\mathcal{D}}_\mathbf{k}(\mathbf{q})$ will become
\begin{align}
  \partial_{\mathbf{q}} \left(\tilde{\mathcal{D}}_\mathbf{k}(\mathbf{q})\tilde{\mathcal{D}}^\dagger_\mathbf{k}(\mathbf{q})\right)\Big{|}_{\mathbf{q} = 0} & = \partial_{\mathbf{q}} \left(\tilde{\mathcal{D}}_\mathbf{k}(\mathbf{q})\right)\Big{|}_{\mathbf{q} = 0} \Delta_1 w^\dagger(\mathbf{k}) + \Delta_1 w(\mathbf{k})\partial_{\mathbf{q}} \left(\tilde{\mathcal{D}}^\dagger_\mathbf{k}(\mathbf{q})\right)\Big{|}_{\mathbf{q} = 0}\nonumber \\
  &= \Delta_1^2 \left( -\partial_\mathbf{k}\tilde{U}^\dagger_\mathbf{k}\tilde{U}_{\mathbf{k}} + \tilde{U}^\dagger_\mathbf{k}\partial_\mathbf{k}\tilde{U}_{\mathbf{k}} + \partial_\mathbf{k}w(\mathbf{k})w^\dagger(\mathbf{k}) -\tilde{U}^\dagger_\mathbf{k}\partial_\mathbf{k}\tilde{U}_{\mathbf{k}} + \partial_\mathbf{k}\tilde{U}^\dagger_\mathbf{k}\tilde{U}_{\mathbf{k}} + w(\mathbf{k})\partial_\mathbf{k} w^\dagger(\mathbf{k}) \right)\nonumber\\
  & = \Delta_1^2 \partial_{\mathbf{k}}\left(w(\mathbf{k})w^\dagger(\mathbf{k})\right)\nonumber\\
  & = 0\,,
\end{align}
\begin{align}
  \partial_{\mathbf{q}} \left(\tilde{\mathcal{D}}^\dagger_\mathbf{k}(\mathbf{q})\tilde{\mathcal{D}}_\mathbf{k}(\mathbf{q})\right)\Big{|}_{\mathbf{q} = 0} =& \partial_{\mathbf{q}} \left(\tilde{\mathcal{D}}^\dagger_\mathbf{k}(\mathbf{q})\right)\Big{|}_{\mathbf{q} = 0} \Delta_1 w(\mathbf{k}) + \Delta_1 w^\dagger(\mathbf{k})\partial_{\mathbf{q}} \left(\tilde{\mathcal{D}}_\mathbf{k}(\mathbf{q})\right)\Big{|}_{\mathbf{q} = 0}\nonumber \\
  =& \Delta_1^2 \left( -w^\dagger(\mathbf{k})\tilde{U}^\dagger_\mathbf{k}\partial_\mathbf{k}\tilde{U}_\mathbf{k}w(\mathbf{k}) + w^\dagger(\mathbf{k})\partial_\mathbf{k}\tilde{U}^\dagger_\mathbf{k}\tilde{U}_{\mathbf{k}} w(\mathbf{k}) + \partial_\mathbf{k}w^\dagger(\mathbf{k})w(\mathbf{k})\right.\nonumber\\ 
  & \left.- w^\dagger(\mathbf{k})\partial_\mathbf{k}\tilde{U}^\dagger_\mathbf{k}\tilde{U}_{\mathbf{k}}w(\mathbf{k}) + w^\dagger(\mathbf{k})\tilde{U}^\dagger_\mathbf{k} \partial_\mathbf{k}\tilde{U}_{\mathbf{k}}w(\mathbf{k}) + w^\dagger(\mathbf{k})\partial_\mathbf{k} w(\mathbf{k}) \right)\nonumber\\
  =& \Delta_1^2 \partial_{\mathbf{k}}\left(w^\dagger(\mathbf{k})w(\mathbf{k})\right)\nonumber\\
  =& 0\,.
\end{align}
Then by using Feynman-Hellman's theorem, we have $\partial_{q_i}\lambda_{\mathbf{k}n}(\mathbf{q})|_{\mathbf{q}=0} = \partial_{q_i}\varphi_{\mathbf{k}n}(\mathbf{q})|_{\mathbf{q}=0} = 0$. Hence the superfluid weight can be further simplified as shown:
\begin{align}
  [D_s]_{ij} &= -\frac{1}{8E_B}\int\frac{d^2k}{(2\pi)^2}{\rm Tr}\left[\partial_{q_i}\partial_{q_j}\left(\tilde{\mathcal{D}}_\mathbf{k}(\mathbf{q})\tilde{\mathcal{D}}^\dagger_\mathbf{k}(\mathbf{q})\right)\Big{|}_{\mathbf{q}=0} + \partial_{q_i}\partial_{q_j}\left(\tilde{\mathcal{D}}^\dagger_\mathbf{k}(\mathbf{q})\tilde{\mathcal{D}}_\mathbf{k}(\mathbf{q})\right)\Big{|}_{\mathbf{q}=0}\right]\nonumber\\
  &= -\frac{1}{8E_B}\int\frac{d^2k}{(2\pi)^2} \partial_{q_i}\partial_{q_j}\left[ {\rm Tr}\left(\tilde{\mathcal{D}}_\mathbf{k}(\mathbf{q})\tilde{\mathcal{D}}^\dagger_\mathbf{k}(\mathbf{q}) + \tilde{\mathcal{D}}^\dagger_\mathbf{k}(\mathbf{q})\tilde{\mathcal{D}}_\mathbf{k}(\mathbf{q})\right) \right]\Big{|}_{\mathbf{q}=0} \nonumber\\
  &= -\frac{1}{4E_B}\int\frac{d^2k}{(2\pi)^2}{\rm Tr}\left[\partial_{q_i}\partial_{q_j}\left(\tilde{\mathcal{D}}_\mathbf{k}(\mathbf{q})\tilde{\mathcal{D}}^\dagger_\mathbf{k}(\mathbf{q})\right)\Big{|}_{\mathbf{q}=0} \right]\nonumber\\
  &= \frac{\Delta_1(\Delta_1 + \Delta_2)}{\sqrt{(\varepsilon_0-\mu)^2 + \Delta_1^2}}\int \frac{d^2k}{(2\pi)^2}{\rm Tr}\left[\frac{1}{2}\left(\partial_{k_i}\tilde{U}^\dagger_\mathbf{k}\partial_{k_j}\tilde{U}_\mathbf{k}+\partial_{k_j}\tilde{U}^\dagger_\mathbf{k}\partial_{k_i}\tilde{U}_\mathbf{k}\right)+ \left( \tilde{U}^\dagger_\mathbf{k}\partial_{k_i}\tilde{U}_\mathbf{k}\tilde{U}^\dagger_\mathbf{k}\partial_{k_j}\tilde{U}_\mathbf{k} \right)\right]\label{eqn:flatbandsw}\,,
\end{align}
where the integrand of this result is the Fubini-Study metric mentioned in the main text, and it will be discussed in detail in Sec. \ref{sec:metric}. Because of the pairing term, the chemical potential is no longer at the flat band energy $\varepsilon_0$. In fact, it can be determined by the filling ratio $\nu$. Bogoliubov theory tells us that the filling ratio of momentum $\mathbf{k}$ is given by $|v_\mathbf{k}|^2$:
\begin{equation}
  v_{\mathbf{k}}^2 = \frac{1}{2}\left(1-\frac{\varepsilon_\mathbf{k}-\mu}{\sqrt{(\varepsilon_\mathbf{k}-\mu)^2 + \Delta_1^2}}\right)\,,
\end{equation}
and in the flat band limit, $\varepsilon_\mathbf{k} = \varepsilon_0$ is momentum independent, which leads to a $\mathbf{k}$ independent $v_\mathbf{k}$. Therefore, the filling ratio of the flat bands is $\nu = v^2_\mathbf{k}$. By using this relation, the superfluid weight at $T=0$ can be written as
\begin{equation}\label{eqn:swflat}
  [D_s]_{ij} = 2|\Delta_1|\left(1 + \frac{\Delta_2}{\Delta_1}\right)\sqrt{\nu(1-\nu)}\int\frac{d^2k}{(2\pi)^2}{\rm Tr}\left[\frac{1}{2}\left(\partial_{k_i}\tilde{U}^\dagger_\mathbf{k}\partial_{k_j}\tilde{U}_\mathbf{k}+\partial_{k_j}\tilde{U}^\dagger_\mathbf{k}\partial_{k_i}\tilde{U}_\mathbf{k}\right)+ \left( \tilde{U}^\dagger_\mathbf{k}\partial_{k_i}\tilde{U}_\mathbf{k}\tilde{U}^\dagger_\mathbf{k}\partial_{k_j}\tilde{U}_\mathbf{k} \right)\right]\,.
\end{equation}
In fact the integrand in Eq. (\ref{eqn:swflat}) is the Fubini-Study metric. It will be discussed in the following section.

\section{quantum geometric tensor, fubini-study metric and berry curvature: lower bounds}\label{sec:metric}
Fubini-Study metric mentioned in last section is closely related with Berry curvature through the ``quantum geometric tensor''. Suppose we have $n$ orthonormal vectors $u_m(\mathbf{k})$, $m=1\ldots n$, in a $N$ dimensional Hilbert space, where $\mathbf{k}$ is some parameter. The quantum geometric tensor can be defined as:
\begin{equation}
  \mathfrak{G}^{mn}_{ij}(\mathbf{k}) = \sum_{a,b=1}^N \partial_{k_i} u^*_{a,m}(\mathbf{k})\left(\delta_{a,b} - \sum_l^n u_{a,l}(\mathbf{k})u^*_{b,l}(\mathbf{k}) \right)\partial_{k_j}u_{b,n}(\mathbf{k})\,,
\end{equation}
in which $m,n$ are energy band indices and $i,j$ are spatial direction indices. For convenience we denote $\tilde{u}_\mathbf{k} = (u_{n_1}(\mathbf{k}), u_{n_2}(\mathbf{k}),\cdots,  u_{n_n}(\mathbf{k}))$ where $n_1,n_2,\cdots, n_n$ are the indices of the energy bands we are interested in. By using $\tilde{u}_\mathbf{k}$, the quantum geometric tensor can be expressed in a more compact form: 
\begin{equation}
  \mathfrak{G}_{ij} = \partial_{k_i}\tilde{u}_\mathbf{k}^\dagger\left(\mathds{1} - \tilde{u}_\mathbf{k}\tilde{u}^\dagger_\mathbf{k}\right)\partial_{k_j}\tilde{u}_\mathbf{k}\,.
\end{equation}
The Hermitian (real) and anti-Hermitian (imaginary) parts of $\mathfrak{G}_{ij}$ are:
\begin{align}
  {\rm Re}[\mathfrak{G}_{ij}] &= \frac{1}{2}\left( \mathfrak{G}_{ij} + \mathfrak{G}_{ij}^\dagger \right)\nonumber\\
  &=\frac{1}{2}\left( \partial_{k_i}\tilde{u}^\dagger_\mathbf{k}\partial_{k_j}\tilde{u}_\mathbf{k} + \partial_{k_j}\tilde{u}^\dagger_\mathbf{k}\partial_{k_i}\tilde{u}_\mathbf{k} + \tilde{u}^\dagger\partial_{k_i}\tilde{u}_\mathbf{k}\tilde{u}^\dagger\partial_{k_j}\tilde{u}_\mathbf{k} + \tilde{u}^\dagger\partial_{k_j}\tilde{u}_\mathbf{k}\tilde{u}^\dagger\partial_{k_i}\tilde{u}_\mathbf{k}\right)\,,\label{eqn:realpartG}\\
  {\rm Im}[\mathfrak{G}_{ij}] &= \frac{1}{2i}\left(  \mathfrak{G}_{ij} - \mathfrak{G}_{ij}^\dagger \right) \nonumber\\
  &= \frac{1}{2i}\left( \partial_{k_i}\tilde{u}^\dagger_\mathbf{k}\partial_{k_j}\tilde{u}_\mathbf{k} - \partial_{k_j}\tilde{u}^\dagger_\mathbf{k}\partial_{k_i}\tilde{u}_\mathbf{k} + \tilde{u}^\dagger\partial_{k_i}\tilde{u}_\mathbf{k}\tilde{u}^\dagger\partial_{k_j}\tilde{u}_\mathbf{k} - \tilde{u}^\dagger\partial_{k_j}\tilde{u}_\mathbf{k}\tilde{u}^\dagger\partial_{k_i}\tilde{u}_\mathbf{k} \right) \,.\label{eqn:imaginarypartG}
\end{align}
For convenience we use $\mathfrak{g}_{ij}$ to denote ${\rm Re}\,\mathfrak{G}_{ij}$, and $g_{ij} = {\rm Tr}\,\mathfrak{g}_{ij}$ is the ``Fubini-Study metric'' mentioned earlier. It is easy to notice that $g_{ij}$ is actually the integrand appeared in Eqs. (\ref{eqn:swflat}). If the Berry connection is defined by $\mathbf{A} = i\tilde{u}^\dagger_\mathbf{k}\partial_{\mathbf{k}}\tilde{u}_\mathbf{k}$, then the imaginary part of $\mathfrak{G}_{ij}$ is proportional to the Berry curvature: ${\rm Im}\,\mathfrak{G}_{ij} = -\frac12 \mathcal{F}_{ij} = -\frac12 (\partial_{k_i}A_j -\partial_{k_j}A_i - i[A_i,A_j])$. So in conclusion, the quantum geometric tensor can be written as
\begin{equation}
  \mathfrak{G}_{ij} = \mathfrak{g}_{ij} -\frac{i}{2}\mathcal{F}_{ij}\,,
\end{equation}

By this quantity, the Fubini-Study metric and Berry curvature are related with each other. An important feature of $\mathfrak{G}_{ij}$ is its positive definiteness. Suppose we have several complex vectors $c_i \in \mathbb{C}^n$, and we will get the following equation:
\begin{align}
  \sum_{ij}c^\dagger_i \mathfrak{G}_{ij}c_j =& \sum_{ij}\sum_{l,m=1}^nc^*_{i,l}\mathfrak{G}^{lm}_{ij}c_{j,m} \nonumber\\
  =&  \left(\sum_i c^\dagger_i \partial_{k_i}\tilde{u}^\dagger_\mathbf{k} \right)\left(\mathds{1}-\tilde{u}_\mathbf{k}\tilde{u}^\dagger_\mathbf{k}\right) \left(\sum_i \partial_{k_i}\tilde{u}_\mathbf{k}c_i \right) \nonumber\\
  =&  \varphi^\dagger \left(\mathds{1}-\tilde{u}_\mathbf{k}\tilde{u}^\dagger_\mathbf{k}\right) \varphi \,\label{eqn:provepositive}\\
  \varphi =&  \sum_i \partial_{k_i}\tilde{u}_\mathbf{k}c_i \,.\nonumber
\end{align}
Because $\{u_m(\mathbf{k})\}$ are orthonormal vectors, the matrix $\left(\mathds{1}-\tilde{u}_\mathbf{k}\tilde{u}^\dagger_\mathbf{k}\right)$ is a projector, and the eigenvalues can only be $0$ or $1$. Therefore a scalar product $\varphi^\dagger\left(\mathds{1}-\tilde{u}_\mathbf{k}\tilde{u}^\dagger_\mathbf{k}\right)\varphi$ must be non-negative. If we choose the complex vectors $c_i$ properly, Eq. (\ref{eqn:provepositive}) can be used to prove important inequalities between the Fubini-Study metric and Berry curvature, as we will show in Sec. \ref{sec:lowerbound}.

The geometric meaning of Fubini-Study metric can be understood as distance between quantum states. In fact the Bloch wave functions of $N$ bands $\tilde{u}_\mathbf{k}$ define a map from the Brillouin zone torus to $\mathbb{CP}^{N-1}$. If we define the distance between two points $\mathbf{k}$ and $\mathbf{k} + d\mathbf{k}$ as shown:
\begin{equation}
  d^2(\mathbf{k},\mathbf{k}+ d\mathbf{k}) = \frac{1}{2}{\rm Tr}\left(\tilde{u}_\mathbf{k}\tilde{u}^\dagger_{\mathbf{k}} -\tilde{u}_{\mathbf{k} + d\mathbf{k}}\tilde{u}^\dagger_{\mathbf{k} + d\mathbf{k}} \right)^2\,,
\end{equation}
then by expanding this equation to the second order we will find $d^2(\mathbf{k},\mathbf{k} + d\mathbf{k}) = g_{ij}(\mathbf{k}_1)dk_idk_j$. A pedagogical introduction to quantum distance and quantum geometric tensor can be found in Ref. \cite{cheng2010quantum}. 

A physical interpretation of the Fubini-Study metric is related with the Wannier function localization which is studied in Ref. \cite{Vanderbilt2012Wannier,Vanderbelt1997Wannier}. If the Bloch wave function of a state in band $n$ and momentum $\mathbf{k}$ is $|\psi_{\mathbf{k}n}$, then the Wannier states can be obtained by the discrete Fourier transformation
\begin{equation}\label{eqn:wannierdef}
  |\mathbf{R}n\rangle = \frac{1}{\sqrt{N}}\sum_\mathbf{k}e^{-i\mathbf{k}\cdot\mathbf{R}}|\psi_{n\mathbf{k}}\rangle
\end{equation}

The Wannier function localization functional can be defined as follows:
\begin{equation}
  F = \sum_n\left[\langle 0 n | \hat{{\mathbf r}}^2 |0n\rangle - |\langle 0 n|\hat{\mathbf{r}}|0n \rangle |^2\right]\,,
\end{equation}
where $\hat{\mathbf{r}}$ is the position operator. However this functional is represented in the Wannier basis. We will show how to express it under Bloch basis. It is well known that the Bloch wave function can be expressed as the product of a periodic function $u_{n\mathbf{k}}$ (the Bloch function) and a phase $e^{i\mathbf{k}\cdot\hat{\mathbf{r}}}$:
$$
|\psi_{n \mathbf{k}} \rangle = e^{i\mathbf{k}\cdot \hat{\mathbf{r}}} | u_{n \mathbf{k}}\rangle\,.
$$
The overlap between two Bloch functions with different momentum will be:
\begin{equation}\label{eqn:overlap}
    \langle u_{m \mathbf{k}}|u_{n \mathbf{k}+ \mathbf{q}}\rangle = \langle \psi_{m\mathbf{k}}| e^{-i\mathbf{q}\cdot \hat{\mathbf{r}}}|\psi_{n{\mathbf{k}+\mathbf{q}}}\rangle\,.
  \end{equation} 
The right hand side of this equation is the Bloch states, and we can transform it into Wannier states as shown:
\begin{equation}
  \langle u_{m \mathbf{k}}|u_{n \mathbf{k}+ \mathbf{q}}\rangle = \frac1N\sum_{\mathbf{R}\mathbf{R}'}e^{-i\mathbf{k}\cdot (\mathbf{R}'-\mathbf{R})}\langle \mathbf{R}' m|e^{-i\mathbf{q}\cdot \hat{\mathbf{r}}}|\mathbf{R}n\rangle e^{-i\mathbf{q}\cdot\mathbf{R}}\,.
\end{equation} 
Because the Wannier functions have a discrete translation symmetry along Bravias lattice, we obtain
$$
\langle \mathbf{R}'m|e^{-i\mathbf{q}\cdot \hat{\mathbf{r}}}|\mathbf{R}n\rangle e^{i\mathbf{q}\cdot \mathbf{R}} = \langle (\mathbf{R}'-\mathbf{R})m|  e^{-i\mathbf{q}\cdot \hat{\mathbf{r}}}| 0n\rangle\,,
$$
 Eq. (\ref{eqn:overlap}) now becomes
$$
  \langle u_{m \mathbf{k}}|u_{n \mathbf{k}+ \mathbf{q}}\rangle = \frac1N\sum_{\mathbf{R}}e^{-i\mathbf{k}\cdot\mathbf{R}}\langle \mathbf{R}m| e^{-i\mathbf{q}\cdot \hat{\mathbf{r}}}|0n\rangle\,.
$$
Up till now we were using the periodic Bloch wave functions $|u_{\mathbf{k}n}\rangle$ instead of the eigenvectors $u_m(\mathbf{k})$ of matrix $\mathcal{H}(\mathbf{k})$. In fact it can be shown that for both continuum models and tight binding models with exactly localized atomic orbitals, we have $\langle u_{m\mathbf{k}}|u_{n\mathbf{k} + \mathbf{q}} \rangle = u^\dagger_m(\mathbf{k})u_n(\mathbf{k} + \mathbf{q}) $. 

The Bloch wave function is given by the following equation in both tight binding models and continuum models:
$$
  |\psi_{m\mathbf{k}}\rangle  = \sum_{\alpha} u_{\alpha,m}(\mathbf{k}) |\mathbf{k},\alpha\rangle\,.
$$
In continuum models, the states $|\mathbf{k},\alpha\rangle$ are given by the plain waves:
\begin{equation}
  \langle \mathbf{r}| \mathbf{k},\alpha \rangle = \frac{1}{\sqrt{N\Omega_c}}e^{i(\mathbf{k} + \mathbf{Q}_\alpha)\cdot \mathbf{r}}\,,
\end{equation}
where $\mathbf{Q}_\alpha$ is a reciprocal vector in momentum space. By using the plain wave basis, we have
\begin{align}
  \langle u_{m\mathbf{k}}|u_{n\mathbf{k} + \mathbf{q}} \rangle &= \sum_{\alpha\beta}u^*_{\alpha,m}(\mathbf{k})u_{\beta,n}(\mathbf{k} + \mathbf{q}) \langle \mathbf{k},\alpha|e^{-i\mathbf{q}\cdot \hat{\mathbf{r}}} |\mathbf{k} + \mathbf{q},\beta\rangle\nonumber\\
  & = \sum_{\alpha\beta}u^*_{\alpha,m}(\mathbf{k})u_{\beta,n}(\mathbf{k} + \mathbf{q})\frac{1}{N\Omega_c} \int d^2r\,e^{-i(\mathbf{k}+\mathbf{Q}_\alpha)\cdot \mathbf{r}}e^{i(\mathbf{k}+\mathbf{q}+\mathbf{Q}_\beta)\cdot\mathbf{r}}e^{-i\mathbf{q}\cdot \mathbf{r}}\nonumber\\
  & = \sum_{\alpha\beta}u^*_{\alpha,m}(\mathbf{k})u_{\beta,n}(\mathbf{k} + \mathbf{q})\frac{1}{N\Omega_c} \int d^2r\,e^{i(\mathbf{Q}_\beta-\mathbf{Q}_\alpha)\cdot \mathbf{r}}\nonumber\\
  &= \sum_{\alpha\beta}u^*_{\alpha,m}(\mathbf{k})u_{\beta,n}(\mathbf{k} + \mathbf{q}) \delta_{\alpha\beta}\nonumber\\
  & = u^\dagger_m(\mathbf{k})u_n(\mathbf{k}+ \mathbf{q})\,.
\end{align}
In tight binding models, the states $|\mathbf{k},\alpha\rangle$ are given by
\begin{equation}
  \langle \mathbf{r}|\mathbf{k},\alpha\rangle =\frac{1}{\sqrt{N}} \sum_{\mathbf{R}}e^{i\mathbf{k}\cdot(\mathbf{R} + \tau_\alpha)} w_\alpha(\mathbf{r}-\mathbf{R}-\tau_\alpha)\,, 
\end{equation}
where $w_\alpha(\mathbf{r}-\mathbf{R}-\tau_\alpha)$ is the local wave function in the unit cell $\mathbf{R}$ and orbital $\alpha$. The inner product of periodic Bloch wave function with different momenta will be
\begin{align}
  \langle u_{m\mathbf{k}}|u_{n\mathbf{k} + \mathbf{q}} \rangle = \sum_{\alpha\beta} u^*_{\alpha,n}(\mathbf{k})u_{\beta,n}(\mathbf{k} + \mathbf{q})\frac{1}{N}\sum_{\mathbf{R}_1,\mathbf{R}_2}e^{-i\mathbf{k}\cdot(\mathbf{R}_1 + \tau_\alpha)}e^{i(\mathbf{k}+\mathbf{q})\cdot(\mathbf{R}_2+\tau_\beta)}\int d^2r\,e^{-i\mathbf{q}\cdot\mathbf{r}}w_\alpha^*(\mathbf{r}-\mathbf{R}_1-\tau_\alpha)w_\beta(\mathbf{r}-\mathbf{R}_2 - \tau_\beta)\,,
\end{align}
If the wave functions of atomic orbitals are assumed to be exactly localized as $w^*_\alpha(\mathbf{r}-\mathbf{R}_1-\tau_\alpha)w_\beta(\mathbf{r}-\mathbf{R}_2-\tau_\beta) = \delta(\mathbf{r}-\mathbf{R}_1-\tau_\alpha)\delta_{\mathbf{R}_1,\mathbf{R}_2}\delta_{\alpha,\beta}$, we will obtain the following equation
\begin{align}
  \langle u_{m\mathbf{k}}|u_{n\mathbf{k} + \mathbf{q}} \rangle &= \sum_{\alpha\beta} u^*_{\alpha,n}(\mathbf{k})u_{\beta,n}(\mathbf{k} + \mathbf{q})\frac{1}{N}\sum_{\mathbf{R}_1,\mathbf{R}_2}e^{-i\mathbf{k}\cdot(\mathbf{R}_1 + \tau_\alpha)}e^{i(\mathbf{k}+\mathbf{q})\cdot(\mathbf{R}_2+\tau_\beta)}e^{-i\mathbf{q}\cdot(\mathbf{R}_1+\tau_\alpha)}\delta_{\alpha,\beta}\delta_{\mathbf{R}_1,\mathbf{R}_2}\nonumber\\
  &= \sum_\alpha u^*_{\alpha,m}(\mathbf{k}) u_{\alpha,n}(\mathbf{k}+\mathbf{q})\frac{1}{N}\sum_{\mathbf{R}}e^{i(-\mathbf{k}+\mathbf{k}+\mathbf{q}-\mathbf{q})\cdot(\mathbf{R}_1 + \tau_\alpha)}\nonumber\\
  & = \sum_{\alpha}u^*_{\alpha,m}(\mathbf{k})u_{\alpha,n}(\mathbf{k} + \mathbf{q})\nonumber\\
  & = u^\dagger_m(\mathbf{k})u_n(\mathbf{k}+ \mathbf{q})\,.
\end{align}
Now we have proved that for both continuum models and tight binding models, the inner product of periodic Bloch wave functions $|u_{n\mathbf{k}}\rangle$ is equal to the inner product of the eigenvectors $u_n(\mathbf{k})$ of $\mathcal{H}(\mathbf{k})$. Therefore, Eq. (\ref{eqn:overlap}) can be written as
\begin{equation}
  u^\dagger_{m}(\mathbf{k})u_n(\mathbf{k}+ \mathbf{q}) = \frac{1}{N}\sum_\mathbf{R}e^{-i\mathbf{k}\cdot \mathbf{R}}\langle\mathbf{R}m|e^{-i\mathbf{q}\cdot \hat{\mathbf{r}}}|0n\rangle\,.
\end{equation}
Then we take the first and second order derivatives of $\mathbf{q}$ on both sides of this equation, and evaluate the result at $q=0$ to obtain:
\begin{align}
  u^\dagger_m(\mathbf{k})\nabla_\mathbf{k}u_n(\mathbf{k}) &= -i\sum_\mathbf{R} e^{-i\mathbf{k}\cdot \mathbf{R}}\langle \mathbf{R}m|\hat{\mathbf{r}}|0n\rangle \label{eqn:rexpectationk}\\
  u^\dagger_m(\mathbf{k})\nabla^2_\mathbf{k}u_n(\mathbf{k}) & = -\sum_\mathbf{R}e^{-i\mathbf{k}\cdot \mathbf{R}}\langle \mathbf{R}m|\hat{\mathbf{r}}^2|0n\rangle \label{eqn:r2expectationk}\,.
\end{align}
The Fourier transformation of these two equations will be
\begin{align}
  \langle \mathbf{R}m|\hat{\mathbf{r}}|0n\rangle &= i\frac{\Omega_c}{(2\pi)^2}\int d^2k\,u^\dagger_m(\mathbf{k})\nabla_\mathbf{k}u_n(\mathbf{k}) e^{i\mathbf{k}\cdot\mathbf{R}}\label{eqn:rexpectation}\\
  \langle \mathbf{R}m|\hat{\mathbf{r}}^2|0n\rangle &= -\frac{\Omega_c}{(2\pi)^2}\int d^2k\, u^\dagger_m(\mathbf{k})\nabla^2_\mathbf{k}u_n(\mathbf{k}) e^{i\mathbf{k}\cdot \mathbf{R}}\,,\label{eqn:r2expectation}
\end{align}
where $\Omega_c$ is the area of the unit cell. For future convenience we divide the localization functional $F$ into the following two parts:
\begin{align}
  F &= \sum_n\left[\langle 0 n | \hat{{\mathbf r}}^2 |0n\rangle - |\langle 0 n|\hat{\mathbf{r}}|0n \rangle |^2\right]\nonumber\\
  & = \sum_n\left[\langle 0n|\hat{\mathbf{r}}^2|0n\rangle - \sum_{\mathbf{R}m}|\langle \mathbf{R} m| \hat{\mathbf{r}} | 0n\rangle |^2\right] + \sum_{n}\sum_{\mathbf{R}m \neq 0n}|\langle \mathbf{R}m|\hat{\mathbf{r}}|0n\rangle |^2  \,,\label{eqn:twoparts}
\end{align}
and let us denote the first and second terms in Eq. (\ref{eqn:twoparts}) by $F_I$ and $\tilde{F}$, respectively. Also $\tilde{F}$ is always positive because of its definition. By using Eqs. (\ref{eqn:rexpectation}) and (\ref{eqn:r2expectation}), the first part $F_I$ can be written as the following form: 
\begin{align}
  F_I &=  \sum_{n}\left[-\frac{\Omega_c}{(2\pi)^2}\int d^2k\,  u^\dagger_n(\mathbf{k})\nabla^2_{\mathbf{k}}u_n(\mathbf{k})- \left(\frac{\Omega_c}{(2\pi)^2}\right)^2\sum_{\mathbf{R}m}\int d^2k_1\int d^2k_2 \,e^{i(\mathbf{k}_1 - \mathbf{k}_2)\cdot \mathbf{R}}\left( u^\dagger_m (\mathbf{k}_1)\nabla_{\mathbf{k}_1} u_n(\mathbf{k}_1)\right)\left(\nabla_{\mathbf{k}_2} u^\dagger_n(\mathbf{k}_2) u_m(\mathbf{k}_2)\right)\right]\nonumber\\
  & = \sum_{n}\left[-\frac{\Omega_c}{(2\pi)^2}\int d^2k\, u^\dagger_n(\mathbf{k})\nabla^2_{\mathbf{k}}u_n(\mathbf{k}) + \frac{\Omega_c}{(2\pi)^2}\sum_{m}\int d^2k \left(u^\dagger_m (\mathbf{k})\nabla_{\mathbf{k}} u_n(\mathbf{k}) \right) \left( u^\dagger_n(\mathbf{k})\nabla_{\mathbf{k}} u_m(\mathbf{k})\right)\right]\,,
\end{align}
The first term is always a real number because it is equal to $\langle 0n|\hat{\mathbf{r}}^2|0n\rangle$. So by taking the complex conjugation we obtain
\begin{equation}
  \frac{\Omega_c}{(2\pi)^2}\int d^2k\,  u^\dagger_n( \mathbf{k})\nabla^2_{\mathbf{k}}u_n(\mathbf{k}) = \frac{\Omega_c}{(2\pi)^2}\int d^2k\, \nabla^2_{\mathbf{k}} u^\dagger_n (\mathbf{k})u_n(\mathbf{k})\,.
\end{equation}
Also from the normalization condition we have $u^\dagger_n(\mathbf{k})u_n(\mathbf{k}) = 1$. The second order derivative tells us $u^\dagger_n(\mathbf{k})\nabla^2_{\mathbf{k}}u_n(\mathbf{k}) + \nabla^2_{\mathbf{k}} u^\dagger_n(\mathbf{k})u_n(\mathbf{k}) + 2 \nabla_\mathbf{k} u^\dagger_n(\mathbf{k})\nabla_\mathbf{k} u_n(\mathbf{k}) = 0$. Consequently we will obtain
\begin{equation}
  -\int d^2k\, u^\dagger_n(\mathbf{k})\nabla^2_{\mathbf{k}}u_n(\mathbf{k}) = -\frac{1}{2}\int d^2k\left( u^\dagger_n( \mathbf{k})\nabla^2_{\mathbf{k}}u_n(\mathbf{k}) + \nabla^2_{\mathbf{k}} u^\dagger_n(\mathbf{k})u_n(\mathbf{k}) \right) = \int d^2k\,\nabla_\mathbf{k}u^\dagger_n(\mathbf{k})\nabla_\mathbf{k} u_n(\mathbf{k})\,.
\end{equation}
We use this expression to replace the first term appeared in $F_I$, and the result is
\begin{align}
  F_I &= \sum_{n}\left[\frac{\Omega_c}{(2\pi)^2}\int d^2k\, \nabla_\mathbf{k}u^\dagger_n(\mathbf{k})\nabla_\mathbf{k} u_n(\mathbf{k}) + \frac{\Omega_c}{(2\pi)^2}\sum_{m}\int d^2k\,u^\dagger_m (\mathbf{k})\nabla_{\mathbf{k}} u_n(\mathbf{k}) u^\dagger_n(\mathbf{k})\nabla_{\mathbf{k}} u_m(\mathbf{k})\right]\nonumber\\
  & = \frac{\Omega_c}{(2\pi)^2}\int d^2k\,{\rm tr}\,g(\mathbf{k})\,.
\end{align}
Notice that the integrand is the trace of the Fubini-Study metric ${\rm tr}\,g$ defined by the eigenvectors $u_n(\mathbf{k})$. Since the metric is invariant under gauge transformation, $F_I$ is gauge invariant. Hence if the integral of ${\rm tr}\,g$ has a nonzero lower bound, the gauge invariant part of Wannier function localization functional will also be bounded. Because $\tilde{F}$ is always positive by definition, the lower bound of the gauge invariant part $F_I$ is also the lower bound of the functional $F$ itself. In the following section we will show some examples with bounded Fubini-Study metric.

\section{bounds on the quantum metric for different systems: SSH chain, Chern insulator and \texorpdfstring{$C_{2z}T$}{C2T} fragile topology}\label{sec:lowerbound}
\subsection{SSH Chain}
Several examples of metric lower bound are studied in this section. We start our discussion with SSH model. SSH model is a one dimensional model with chiral symmetry and it can be classified by a winding number. In this paragraph we show that the integral of Fubini-Study metric of SSH chain is bounded by the winding number. If the chiral symmetry is represented by $\sigma_z$, then the (flat band) Hamiltonian of SSH model can be written as
\begin{equation}
  \mathcal{H}(k) = \cos(\phi(k))\sigma_x + \sin(\phi(k))\sigma_y\,,
\end{equation}
in which $\phi(k)$ is a function of $k$. The topological index of this model is the winding number, defined by $\mathcal{W} = \frac{1}{2\pi}\int dk\,\partial_k\phi(k)\in\mathbb{Z}$. By diagonalizing $\mathcal{H}(k)$ and finding the wave function of valence band states, we can find the Fubini-Study metric of SSH model is:
\begin{equation}
  g(k) = \frac{1}{4}\left(\partial_k\phi(k)\right)^2\,,
\end{equation}
then it is easy to show that the integral of metric is bounded by the winding number:
\begin{equation}
  \frac{1}{2\pi}\int dk\,g(k) = \frac{1}{4}\frac{1}{2\pi}\int dk\,(\partial_k\phi(k))^2 \geq \frac{a}{4}\left(\frac{1}{2\pi}\int dk\,\partial_k\phi(k)\right)^2 = \frac{a\mathcal{W}^2}{4}\,,
\end{equation}
in which $a$ is the lattice constant. Hence the integral of Fubini-Study metric in one dimension with chiral symmetry is bounded by the winding number. 

\subsection{Chern Insulator}
The next example will be a band with nonzero Chern number. The integral of Berry curvature of a Chern band must be $2\pi n, n\in\mathbb{Z}$. We consider the case with just one band with a Chern number $C_1$. That means $\mathfrak{G}_{ij} = g_{ij} - \frac{i}{2}\mathcal{F}_{xy}$ is a complex number instead of a matrix and $\mathfrak{g}_{ij} = g_{ij}$. If we choose $c_x = 1, c_y = i$, Eq. (\ref{eqn:provepositive}) will become:
\begin{equation}
  \mathfrak{G}_{xx} + \mathfrak{G}_{yy} + i\mathfrak{G}_{xy} - i\mathfrak{G}_{yx} = g_{xx} + g_{yy} +\mathcal{F}_{xy} \geq 0~~~\Rightarrow~~~{\rm tr}\,g\geq -\mathcal{F}_{xy}\,.
\end{equation}
Similarly, if we choose $c_x = 1, c_y = -i$, we will get ${\rm tr}\,g\geq \mathcal{F}_{xy}$. In conclusion, the trace of the metric is bounded by the local Berry curvature ${\rm tr}\, g \geq |\mathcal{F}_{xy}|$. Obviously, lower bound for the integral of metric is given by the Chern number:
\begin{equation}
  \frac{1}{2\pi}\int d^2k\,{\rm tr}\,g \geq \Bigg{|} \frac{1}{2\pi} \int d^2k\,\mathcal{F}_{xy} \Bigg{|} = |C_1|\,.
\end{equation}

\subsection{Topology of TBLG flat bands}
The flat bands in TBLG are not Chern bands. In fact, their Chern number vanishes. The effective continuum model of TBLG per spin per valley is the Bistritzer-MacDonald model \cite{bistritzer2011} whose two flat bands around the charge neutral point have a nontrivial topology protected by this $C_{2z}T$ symmetry. The free fermion Hamiltonian of valley $K$ and spin up is given by:
\begin{equation}\label{eqn:moireband}
  h(\mathbf{k})_{\mathbf{Q},\mathbf{Q}'} = v_F\sigma \cdot (\mathbf{k} - \mathbf{Q}) \delta_{\mathbf{Q},\mathbf{Q}'} + w\sum_{j = 1}^3 \left(\delta_{\mathbf{Q}-\mathbf{Q}',\mathbf{q}_j}T_j + \delta_{\mathbf{Q}-\mathbf{Q}',-\mathbf{q}_j}T_j^\dagger \right)\,,
\end{equation} 
in which $\mathbf{Q},\mathbf{Q}'$ take their value in the hexagonal lattice formed by adding $\mathbf{q}_{1,2,3}$ iteratively. Notice that this Hamiltonian is not periodic in momentum space. In fact it changes as $h(\mathbf{k} + \mathbf{G}) = V_\mathbf{G}h(\mathbf{k})V^\dagger_{\mathbf{G}}$ under momentum translation, in which $(V_{\mathbf{G}})_{\mathbf{Q},\mathbf{Q}'}=e^{i\alpha}\delta_{\mathbf{Q}'-\mathbf{Q},\mathbf{G}}$. The choice of $V_\mathbf{G}$ has a phase degree of freedom, and we choose it to be real ($\alpha = 0,\pi$). The reason will be explained later. The $C_{2z} T$ transformation and its representation unitary matrix $D(C_{2z}T)$ are defined as follows:
\begin{equation}\label{eqn:c2tdef}
  (C_{2z}T)\psi_{K\uparrow,\mathbf{Q},a,\mathbf{k}}(C_{2z}T)^{-1} = D(C_{2z}T)_{\mathbf{Q},a;\mathbf{Q}',b}\psi_{K\uparrow,\mathbf{Q}',b,\mathbf{k}}\,,~~~D(C_{2z}T)_{\mathbf{Q},a;\mathbf{Q}',b} = \delta_{\mathbf{Q},\mathbf{Q}'}(\sigma_x)_{ab}\,,
\end{equation}
and the Hamiltonian satisfies $D(C_{2z}T)h^*(\mathbf{k})D^{-1}(C_{2z}T)=h(\mathbf{k})$.

\subsubsection{\texorpdfstring{$C_{2z}T$}{C2T} Sewing matrix and off-diagonal Berry connection}

The topology of these bands has been studied in previous research where it has been proved that they have a Wilson loop winding number \cite{songz2018,ahn2018}. Before we prove that the metric is bounded by that number, we briefly review the topology of these flat bands. In momentum space, $T$ sends $\mathbf{k}$ to $-\mathbf{k}$ and then $C_{2z}$ sends it back to $\mathbf{k}$. Since $T^2 = 1$ in this case, there is no Kramers degeneracy, and we can have the sewing matrix of $C_{2z}T$ symmetry. If $u_n(\mathbf{k})$ are the eigenstates of the Hamiltonian $h(\mathbf{k})$, the $C_{2z}T$ sewing matrix $B(\mathbf{k})$ will have the following form:
\begin{equation}\label{eqn:sewing}
  D(C_{2z} T) u^*_m(\mathbf{k}) = \sum_m u_m(\mathbf{k}) B_{mn}(\mathbf{k})\,,~~~~B(\mathbf{k}) = \tilde{u}^\dagger(\mathbf{k})D(C_{2z} T)\tilde{u}^*(\mathbf{k})\,,
\end{equation}
where $\tilde{u}_\mathbf{k} = (u_1(\mathbf{k}),u_2(\mathbf{k}))$ are the eigenvectors of $h(\mathbf{k})$ which correspond to the flat bands. $C_{2z}T$ symmetry is antiunitary, hence we need a real embedding matrix $V_\mathbf{G}$ to make $B(\mathbf{l})$ periodic:
\begin{equation}
  B(\mathbf{k} + \mathbf{G}) = \tilde{u}^\dagger_{\mathbf{k}+\mathbf{G}}D(C_{2z}T)\tilde{u}^*_{\mathbf{k} + \mathbf{G}} = \tilde{u}^\dagger_{\mathbf{k}}V^\dagger_\mathbf{G}D(C_{2z}T) V^*_\mathbf{G} \tilde{u}^*_{\mathbf{k}} = \tilde{u}^\dagger_{\mathbf{k}}D(C_{2z}T)\tilde{u}^*_{\mathbf{k}} =B(\mathbf{k})\,.
\end{equation}

Because $(C_{2z}T)^2 = 1$, there is no Kramers theorem, then if the two flat bands are not degenerate (away from the Dirac points), the sewing matrix must be diagonal and can be written as 
$$
B(\mathbf{k}) = \left(\begin{array}{cc} e^{i\theta_{1\mathbf{k}}} & 0 \\0 & e^{i\theta_{2\mathbf{k}}} \end{array}\right)\,.
$$
Because of the periodicity of $B(\mathbf{k})$, we must have $(\theta_{1\mathbf{k}+\mathbf{G}} - \theta_{1\mathbf{k}}) = 2\pi n,(\theta_{2\mathbf{k}+\mathbf{G}} - \theta_{2\mathbf{k}}) = 2\pi m$ in which $n,m\in\mathbb{Z}$. By definition, the non-Abelian Berry connection is given by $\mathbf{A}(\mathbf{k}) = i\tilde{u}^\dagger_\mathbf{k}\partial_\mathbf{k}\tilde{u}_\mathbf{k}$, where $\tilde{u}_\mathbf{k}=(u_1(\mathbf{k}),u_2(\mathbf{k}))$ are the eigenvectors of $h(\mathbf{k})$ which correspond to the flat bands. Due to the property of the sewing matrix Eq. (\ref{eqn:sewing}), when $\mathbf{k}$ is not on any Dirac points, the Berry connection satisfies the following equation \cite{songz2018,ahn2018}:
\begin{equation}
  \mathbf{A}(\mathbf{k}) = -B(\mathbf{k})\mathbf{A}^{\rm T}(\mathbf{k})B^\dagger(\mathbf{k}) + iB(\mathbf{k})\partial_\mathbf{k}B^\dagger(\mathbf{k})\,,
\end{equation}
and this equation is a constraint on the matrix elements of the Berry connection. It requires the non-Abelian connection to have the following form:
\begin{equation}\label{eqn:berryconnection}
  \mathbf{A}(\mathbf{k}) =\left(\begin{array}{cc}
    \frac{1}{2}\partial_\mathbf{k} \theta_{1\mathbf{k}} & i \mathbf{a}(\mathbf{k})e^{\frac{i}{2}(\theta_{1\mathbf{k}}-\theta_{2\mathbf{k}})}\\
    -i \mathbf{a}(\mathbf{k})e^{\frac{i}{2}(\theta_{2\mathbf{k}}-\theta_{1\mathbf{k}})} & \frac{1}{2}\partial_\mathbf{k}\theta_{2\mathbf{k}}
  \end{array}
  \right)\,,
\end{equation}
in which $\mathbf{a}(\mathbf{k})$ is a real vector. It is important to know whether $\mathbf{a}(\mathbf{k})$ is single valued or not. Because the embedding matrix $V_\mathbf{G}$ is momentum independent, the Berry connection itself is periodic. Also the winding numbers $n_{1,2}$ defined by phase factors $\theta_{1,2\mathbf{k}+\mathbf{G}} - \theta_{1,2\mathbf{k}} = 2\pi n_{1,2}$ must be integers because of periodicity of $B(\mathbf{k})$. Therefore we have $e^{\frac{i}{2}(\theta_{1\mathbf{k}+\mathbf{G}}-\theta_{2\mathbf{k}+\mathbf{G}})} = \pm e^{\frac{i}{2}(\theta_{1\mathbf{k}}-\theta_{2\mathbf{k}})}$. This means the vector $\mathbf{a}(\mathbf{k})$ can be periodic ($\mathbf{a}(\mathbf{k} + \mathbf{G}) = \mathbf{a}(\mathbf{k})$) if $(n_1 - n_2)$ is an even integer, or antiperiodic ($\mathbf{a}(\mathbf{k} + \mathbf{G}) = -\mathbf{a}(\mathbf{k})$) if $(n_1 - n_2)$ is an odd integer. . In fact, by using the property of $C_{2z}T$ symmetry, we can prove that $\mathbf{a}(\mathbf{k})$ must be periodic for topological nontrivial bands.

When $\mathbf{k}$ is not on any Dirac points, we can write down the form of non-Abelian Berry curvature by Eq. (\ref{eqn:berryconnection}):
\begin{equation}
  \mathcal{F}_{xy} = \partial_{k_x}A_y(\mathbf{k}) - \partial_{k_y}A_x(\mathbf{\mathbf{k}}) - i[A_x(\mathbf{k}),A_y(\mathbf{k})] = \left(\begin{array}{cc}
      0 & if_{xy}e^{\frac{i}{2}(\theta_{1\mathbf{k}}-\theta_{2\mathbf{k}})}\\
      -if_{xy}e^{\frac{i}{2}(\theta_{2\mathbf{k}}-\theta_{1\mathbf{k}})} & 0
    \end{array}
  \right)\,,~~~f_{xy} = \partial_{k_x} a_y - \partial_{k_y} a_x\,.
\end{equation}
We will show how the Wilson loop winding number is related with $f_{xy}$ in the following subsection. 

%%%% Wilson Loop discussion
\subsubsection{Wilson loop winding}

It can be shown that the winding number of the Wilson loop of the two active bands can be expressed as the integral of $f_{xy}$ on the whole Brillouin zone. We can label the momentum as $\mathbf{k} = \frac{k_1}{2\pi}\mathbf{b}_1 + \frac{k_2}{2\pi}\mathbf{b}_2$ in which $\mathbf{b}_{1,2}$ are the primitive vectors of the Moir\'e reciprocal lattice. The large Wilson loop across the whole BZ along $\mathbf{b}_2$ direction with fixed $k_1$ is defined as
\begin{align}
  W(k_1) &= \mathcal{P}\exp\left(-i\oint_c d\mathbf{k}\cdot \mathbf{A}(\mathbf{k})\right)  \nonumber\\
   &= \lim_{\delta k \rightarrow 0}\tilde{u}^\dagger(k_1,0)\tilde{u}(k_1,\delta k)\tilde{u}^\dagger(k_1,\delta k)\tilde{u}(k_1,2\delta k)\tilde{u}^\dagger(k_1,2\delta k)\cdots \tilde{u}^\dagger(k_1,2\pi -\delta k)\tilde{u}(k_1,2\pi)\,,
\end{align}
where $\mathcal{P}$ stands for the path ordering and $c$ is the straight line path from $(k_1,0)$ to $(k_1,2\pi)$. The complex conjugation of this Wilson loop operator will become:
\begin{align}
  W^*(k_1) &= \lim_{\delta k \rightarrow 0}\tilde{u}^{\rm T}(k_1,0)\tilde{u}^*(k_1,\delta k)\tilde{u}^{\rm T}(k_1,\delta k)\tilde{u}^*(k_1,2\delta k)\tilde{u}^{\rm T}(k_1,2\delta k)\cdots \tilde{u}^{\rm T}(k_1,2\pi -\delta k)\tilde{u}^*(k_1,2\pi)\nonumber\\
  &= \lim_{\delta k \rightarrow 0}\tilde{u}^{\rm T}(k_1,0)D^{-1}(C_{2z}T)D(C_{2z}T)\tilde{u}^*(k_1,\delta k)\tilde{u}^{\rm T}(k_1,\delta k)D^{-1}(C_{2z}T)\cdots \tilde{u}^{\rm T}(k_1,2\pi -\delta k)D^{-1}(C_{2z}T)D(C_{2z} T)\tilde{u}^*(k_1,2\pi)\nonumber\\
  &= \lim_{\delta k \rightarrow 0}B^\dagger(k_1,0)\tilde{u}^\dagger(k_1,0)\tilde{u}(k_1,\delta k)B(k_1,\delta k)B^\dagger(k_1,\delta k)\tilde{u}^\dagger(k_1,\delta k) \cdots B^\dagger(k_1,2\pi - \delta k) \tilde{u}^\dagger(k_1 ,2\pi -\delta k) \tilde{u}(k_1,2\pi) B(k_1,2\pi)\nonumber\\
  &= B^\dagger(k_1,0)W(k_1)B(k_1,2\pi)\,.
\end{align}
Since we have proved that the sewing matrix is periodic, we find that $W(k_1)$ and $W^*(k_1)$ only differ by a unitary transformation. There will be two kinds of possiblities: 1) the two eigenvalues are complex conjugation to each other; 2) the two eigenvalues are real, and because of unitarity of $W(k_1)$, the two eigenvalues can only be $\{1,1\}$, $\{1,-1\}$ or $\{-1,-1\}$. However, the second case can only give us a trivial band because the Wilson loop is not winding, and hence we only consider the first case in the following discussion. Because the two eigenvalues are complex conjugation to each other, the determinant of this large Wilson loop must be $1$. Therefore, we have
\begin{equation}\label{eqn:detW}
  {\rm det}\,W(k_1) = \exp\left(-i\oint_c d\mathbf{k}\cdot{\rm Tr}\,\mathbf{A}(\mathbf{k})\right) = 1\,. 
\end{equation}
When there is no Dirac point along this Wilson loop, from Eq. (\ref{eqn:berryconnection}) we know that ${\rm Tr}\,\mathbf{A}(\mathbf{k}) = \frac{1}{2}\partial_\mathbf{k}(\theta_{1\mathbf{k}} + \theta_{2\mathbf{k}})$, thus $\oint_c d\mathbf{k}\cdot {\rm Tr}\,\mathbf{A}(\mathbf{k}) = \pi(n_1+n_2)$, where $n_{1,2}$ are the winding numbers of $\theta_{1,2\mathbf{k}}$ along this large Wilson loop. Because of Eq. (\ref{eqn:detW}), $(n_1 + n_2)$ must be an even integer, and thus $(n_1 - n_2)$ is also an even integer. As we mentioned in last subsection, the off-diagonal element of Berry connection $i\mathbf{a}(\mathbf{k})e^{\frac{i}{2}(\theta_{1\mathbf{k}}-\theta_{2\mathbf{k}})}$ is periodic. Since $e^{\frac{i}{2}(\theta_{1\mathbf{k}+\mathbf{G}}-\theta_{2\mathbf{k}+\mathbf{G}})} = e^{i\pi(n_1 - n_2)} e^{\frac{i}{2}(\theta_{1\mathbf{k}}-\theta_{2\mathbf{k}})} = e^{\frac{i}{2}(\theta_{1\mathbf{k}}-\theta_{2\mathbf{k}})}$, we find that $\mathbf{a}(\mathbf{k} + \mathbf{G}) = \mathbf{a}(\mathbf{k})$.

The above discussion is about a large Wilson loop $W(k_1)$ along BZ without going across any Dirac points. However the topology of bands is usually diagnosed by the winding of Wilson loop over the whole BZ, which means $k_1$ varies from $0$ to $2\pi$ and it must encounter Dirac points during this process. It can be shown that $W(k_1)$ is continuous when it goes across a Dirac point. Assume that $W_D$ is a non-Abelian Wilson loop surrounding a Dirac cone at position $(k_{01},k_{02})$. If we want to shrink this loop to a point with the Dirac point surrounded by it, then the shrinking point must be that Dirac point. The Wilson loop is made out of the projector into the two bands that have the Dirac point, hence it must be the identity. As shown in FIG. \ref{fig:wilsonloop} (a), the Wilson loop $W(k_{01} - \delta)$ and $W(k_{01}+\delta)$ can be represented as 
\begin{align*}
  W(k_{01}-\delta) &= W_1^- W^{-1}_{Dl}W_2^-\,,\\
  W(k_{01}+\delta) &= W_1^+ W_{Dr}W_2^+\,,
\end{align*}
and these Wilson line operators $W_1^\pm$, $W_2^\pm$, $W_{Dl}$ and $W_{Dr}$ are also shown in FIG. \ref{fig:wilsonloop} (a). As we just mentioned, the Wilson loop around the Dirac point is the identity when the radius $D$ of that Wilson loop approaches zero, which means $\mathds{1} = W_{Dl}W_{Dr}$ and hence $ W_{Dr}= W_{Dl}^{-1}$. Also because $W_{1,2}^-$ can be continuously deformed to $W_{1,2}^+$ without crossing with any Dirac point, we will obtain that $W_{1,2}^ - = W_{1,2}^+$ when $\delta\rightarrow 0$. One hence easily finds that $W(k_{01} -\delta)$ and $W(k_{01} +\delta)$ are identical at the limit when both $D$ and $\delta$ go to zero. That is our statement that $W(k_1)$ is continuous when it goes across a Dirac point.

\begin{figure}[!htbp]
\centering
\includegraphics[width=10cm]{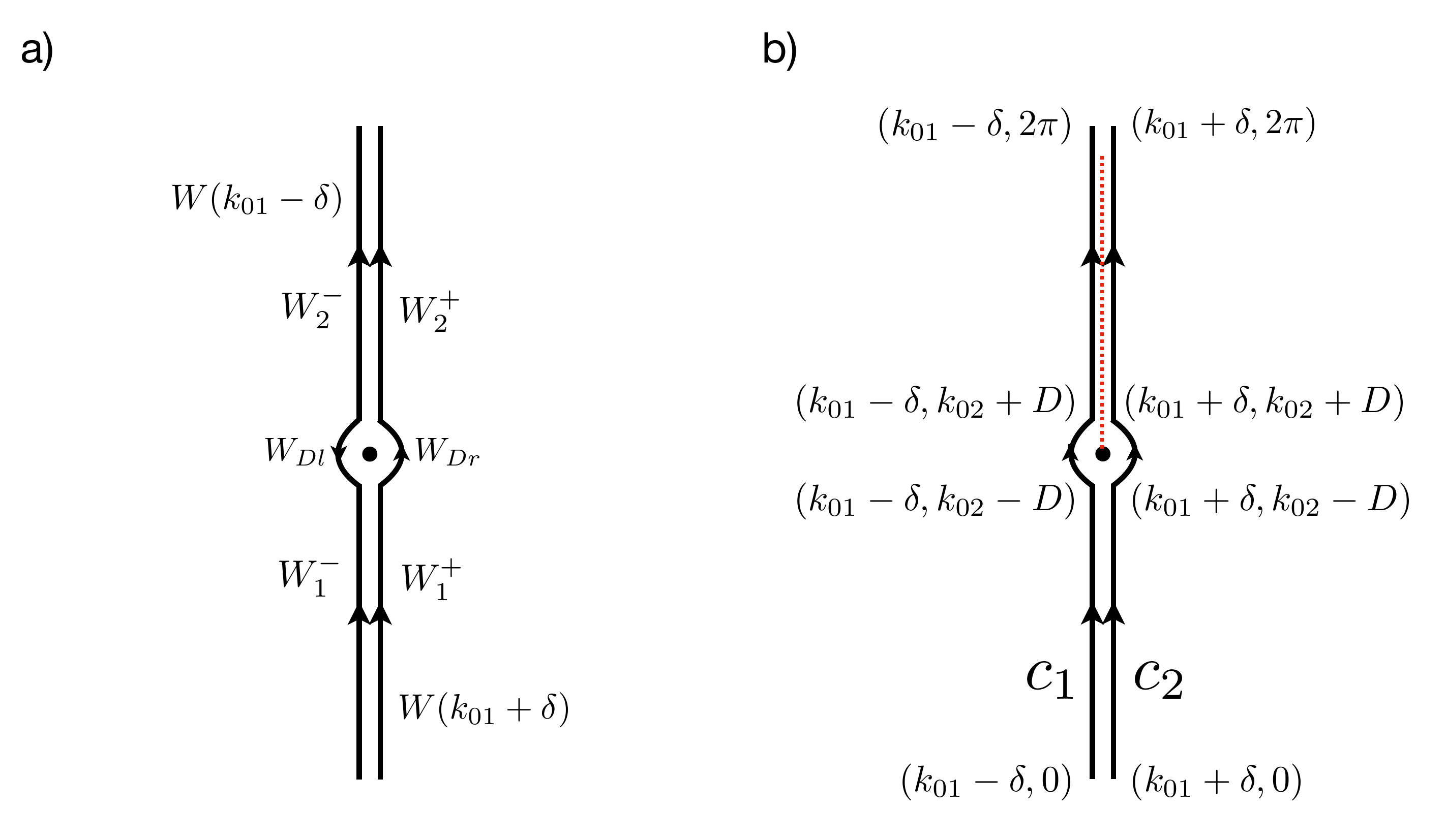}
\caption{a) The Wilson loop $W(k_1)$ is continuum when it goes across a Dirac point. $W_1$, $W_2$ and $W_{Dl(r)}$ are Wilson line operators. b) The winding number of $\theta_{1,2\mathbf{k}}$ changes from even(odd) to odd(even) when it goes across a Dirac point. The red dashed line is the branch cut of the phase factor $\theta_{1,2\mathbf{k}}$.}
\label{fig:wilsonloop}
\end{figure}

Now we study how the winding number of Wilson loop is related with the off-diagonal Berry connection/curvature which defines the Euler class. To do this we need to know the eigenvalue of the Wilson loop operator. However, the form of its eigenvalue depends on the winding numbers of $\theta_{1\mathbf{k}}$ and $\theta_{2\mathbf{k}}$, \emph{i.e.,} $n_1,~n_2$, along the path of that Wilson loop. As we proved in previous paragraph, $(n_1-n_2)$ is an even integer and there will be two possibilities: 1) both of them are even integers, or 2) both of them are odd integers. If we calculate the \emph{Abelian} Wilson loop of one band (either of the bands forming the Dirac node) around a Dirac point, the Berry phase must be $\pm \pi$. This means the loop integral of both $\mathbf{A}_{11}(\mathbf{k})$ or $\mathbf{A}_{22}(\mathbf{k})$ around the Dirac point is either $\pm\pi$.  From Eq. (\ref{eqn:berryconnection}) we know $\mathbf{A}_{11}(\mathbf{k}) = \frac12 \partial_\mathbf{k}\theta_{1\mathbf{k}}$. Thus the winding number of $\theta_{1\mathbf{k}}$ around the Dirac point will be $\pm 2\pi$. Because there is a nonzero winding of $\theta_{1\mathbf{k}}$ around the Dirac point, there will be a branch cut, and the value of $\theta_1$ on two sides of the branch cut differs by $\pm 2\pi$. Now we consider two paths $c_1$ and $c_2$ shown in FIG. \ref{fig:wilsonloop} (b) and the dashed line stands for the branch cut. Both $c_1$ and $c_2$ are a large loop which go across the whole BZ. There is a branch cut between the points $(k_{01}-\delta,2\pi)$ and $(k_{01}+\delta,2\pi)$ , but there is no branch between $(k_{01}-\delta,0)$ and $(k_{01}+\delta,0)$. Hence when $\delta\rightarrow 0$, we have $\theta_1(k_{01}-\delta,0) = \theta_1(k_{01}+\delta,0)$ and $\theta_1(k_{01}-\delta,2\pi) = \theta_1(k_{01}+\delta,2\pi)\pm2\pi$. Therefore, the winding number $n_1$ changes by one when the path goes across a Dirac point. Similar argument can also be applied to $\theta_2$, because $\mathbf{A}_{22}(\mathbf{k}) = \frac12 \partial_\mathbf{k}\theta_{2\mathbf{k}}$. In conclusion, $n_1$ and $n_2$ changes from even (odd) to odd (even) when the path of the Wilson loop goes across a Dirac point.

The eigenvalue of the Wilson loops $W(k_1)$ can be calculated as shown. To obtain an expression without $\theta_{1,2\mathbf{k}}$, we do the following transformation $u'_\mathbf{k} = \tilde{u}_\mathbf{k}g_\mathbf{k}$, in which the transformation $g_\mathbf{k}$ is given by
\begin{equation}\label{eqn:localgauge}
    g_\mathbf{k} = \left(\begin{array}{cc}
      e^{\frac{i}{2}\theta_{1\mathbf{k}}} & 0 \\ 0 & e^{\frac{i}{2}\theta_{2\mathbf{k}}}
    \end{array}
  \right)\,.
\end{equation}
Then the Wilson loop $W(k_1)$ can be expressed by $u'_\mathbf{k}$ as follows:
\begin{align}\label{eqn:wilsonlooplocalgauge}
  W(k_1) &= \lim_{\delta k \rightarrow 0}\tilde{u}^\dagger(k_1,0)\tilde{u}(k_1,\delta k)\tilde{u}^\dagger(k_1,\delta k)\tilde{u}(k_1,2\delta k)\tilde{u}^\dagger(k_1,2\delta k)\cdots \tilde{u}^\dagger(k_1,2\pi -\delta k)\tilde{u}(k_1,2\pi)\nonumber\\
  &= \lim_{\delta k \rightarrow 0} g(k_1,0)u'^\dagger(k_1,0)u'(k_1 ,\delta k)g^\dagger(k_1,\delta k)g(k_1,\delta k)u'^\dagger(k_1,\delta k)\cdots u'(k_1,2\pi)g^\dagger(k_1,2\pi)\nonumber \\
  &= g(k_1,0)W'(k_1)g^\dagger(k_1,2\pi)\,.
\end{align}
When $\theta_{1\mathbf{k}}$ and $\theta_{2\mathbf{k}}$ wind even times, $g(k_1,0) = g(k_1,2\pi)$, the eigenvalues of $W(k_1)$ is the same as $W'(k_1)$. If the winding number is odd, then $g(k_1,0) = -g(k_1, 2\pi)$, the eigenvalue of $W(k_1)$ is the same as $-W'(k_1)$. It is hence important  to know the expression of $W'(k_1)$. From the expression of $W'(k_1)$, we find that it can be written as
\begin{equation}\label{eqn:newwilson}
  W'(k_1) = \mathcal{P}\exp\left(-i\int_0^{2\pi}dk_2\, A'_2(k_1,k_2)\right)\,,~~~\mathbf{A}'(\mathbf{k}) = i u'^\dagger_\mathbf{k} \partial_{\mathbf{k}} u'_\mathbf{k} = g_\mathbf{k}\mathbf{A}(\mathbf{k})g^\dagger_\mathbf{k} + ig_\mathbf{k} \partial_\mathbf{k} g^\dagger_\mathbf{k}\,,
\end{equation}
where $\mathcal{P}$ stands for path ordering. We use Eqs. (\ref{eqn:berryconnection}) and (\ref{eqn:localgauge}) to derive $\mathbf{A}'$, and the result is $\mathbf{A}'(\mathbf{k}) = -\mathbf{a}(\mathbf{k})\sigma_2$. Because $\sigma_2$ always commute with itself, $W'(k_1)$ can be expressed without the use of the path ordering:
\begin{equation}
  W'(k_1) = \exp\left(i\sigma_2 \int_0^{2\pi}dk_2\, a_2(k_1,k_2) \right)\,.
\end{equation}
By this equation we can get the eigenvalue of Wilson loop $W(k_1)$ as shown:
\begin{equation}
  \xi(k_1) = \left\{\begin{array}{ll}
    \int_0^{2\pi}dk_2\,a_2(k_1,k_2)~{\rm mod}~2\pi &  \theta_{1,2}\,\text{wind even times,}\\
    \pi+\int_0^{2\pi}dk_2\,a_2(k_1,k_2)~{\rm mod}~2\pi &  \theta_{1,2}\,\text{wind odd times.}
  \end{array}
  \right.
\end{equation}
At first glance the Wilson loop spectrum is not a continuous function of $k_1$. However, as we will see in the following paragraph, the winding of $\mathbf{a}(\mathbf{k})$ around the Dirac point must be an odd integer, which means $\int_0^{2\pi}dk_2\,a_2(k_1,k_2)$ is not continuous, and as a result we obtain a continuous eigenvalue $\xi(k_1)$. This statement can be proved as follows. By using the transformation in Eq. (\ref{eqn:wilsonlooplocalgauge}), the Wilson loop of the two bands around the Dirac point that they form can be written as
\begin{equation}
  W_D = g_{\mathbf{k}_i} \exp\left(i\sigma_2\oint d\mathbf{k} \cdot \mathbf{a}(\mathbf{k})\right)g^\dagger_{\mathbf{k}_f}\,,
\end{equation}
In this equation the Wilson loop starts at point $\mathbf{k}_i$, and the it goes around the Dirac point and then ends at $\mathbf{k}_f$. Actually $\mathbf{k}_i$ and $\mathbf{k}_f$ are at the same position in the momentum space but the matrix $g_{\mathbf{k}}$ are different because of the winding of $\theta_{1,2}$.  As we proved in the text, the winding numbers of $\theta_{1\mathbf{k}}$ and $\theta_{2\mathbf{k}}$ around the Dirac point are $\pm 2\pi$, which means $g_{\mathbf{k}_i} = -g_{\mathbf{k}_f}$. Also the non-Abelian Wilson loop is identity when the radius of the loop goes to zero $W_D|_{D\rightarrow 0} =\mathds{1}$ , so the following equation holds:
\begin{equation}
  \exp\left(i\sigma_2\oint_D d\mathbf{k} \cdot \mathbf{a}(\mathbf{k})\right) = -\mathds{1}\,,
\end{equation}
where $D$ is the infinitesimal loop around a Dirac point. This means the winding of $\mathbf{a}(\mathbf{k})$ is an odd number times $\pi$:
\begin{equation}
  \oint_D d\mathbf{k}\cdot \mathbf{a}(\mathbf{k}) = (2n + 1) \pi\,,~~~n\in\mathbb{Z}\,.
\end{equation}
Now if a Dirac point is assumed to be located at $(k_{01},k_{02})$, we find that the integral $\int_0^{2\pi}dk_2\,a_2(k_1,k_2)$ is not continuous and it satisfies:
\begin{equation}\label{eqn:discontinuousa}
  \lim_{\delta\rightarrow 0}\left(\int_0^{2\pi}dk_2\,a_2(k_{01}+\delta,k_2) - \int_0^{2\pi}dk_2\,a_2(k_{01}-\delta,k_2)\right)= \lim_{\delta\rightarrow 0,D\rightarrow 0}\left(\int_{c_1^++c_2^+ + c_3^+}d\mathbf{k}\cdot \mathbf{a}(\mathbf{k})-\int_{c_1^-+c_2^- + c_3^-}d\mathbf{k}\cdot \mathbf{a}(\mathbf{k})\right)\,,
\end{equation}
where the paths $c_{1,2,3}^\pm$ are defined in FIG. \ref{fig:windingstokes} (a). When $\delta\rightarrow 0$, we have $\int_{c_{1,3}^+}d\mathbf{k}\cdot \mathbf{a}(\mathbf{k}) = \int_{c_{1,3}^-}d\mathbf{k}\cdot \mathbf{a}(\mathbf{k})$. Therefore, Eq. (\ref{eqn:discontinuousa}) becomes the integral of $\mathbf{a}(\mathbf{k})$ along the path $c_2^+ - c_2^-$, which is actually a loop around the Dirac point. This can tell us that $\int_0^{2\pi}dk_2\,a_2(k_1,k_2)$ jumps by $(2n + 1) \pi$ at $k_{10}$ if a Dirac point is located at $(k_{01},k_{02})$. Hence, the eigenvalue of $W(k_1)$ is continuous, because
\begin{equation}\label{eqn:stokes}
  \lim_{\delta\rightarrow 0}(\xi(k_{01}+\delta)-\xi(k_{01} - \delta)) = 0~{\rm mod}~2\pi\,.
\end{equation}
That result is consistent with our argument shown in FIG. \ref{fig:wilsonloop} (a). Now suppose there is no Dirac point between $k_1$ and $k_1'$, then by using Stokes' theorem, the increment of $\xi$ is given by the integral of $f_{xy}$
\begin{equation}
  \xi(k_1') - \xi(k_1) = \int_{\Omega_{k_1,k_1'}} d^2k \,f_{xy}\,,
\end{equation}
where $\Omega_{k_1,k_1'}$ is a rectangle with points $(k_1,0),~(k_1,2\pi),~(k_1',2\pi)$ and $(k_1',0)$. If there is a Dirac point between $k_1$ and $k_1'$ at $k_{01}$, then we can apply Stokes' theorem on both sides of Dirac point as shown in FIG. \ref{fig:windingstokes} (b), and we obtain the following equations
\begin{equation}
  \xi(k_1') - \xi(k_{01}+\delta) = \int_{\Omega_{k_{01}+\delta,k_1'}} d^2k\,f_{xy}\,,~~~\xi(k_{01}-\delta) - \xi(k_1) = \int_{\Omega_{k_1,k_{01}-\delta}} d^2 k\,f_{xy}\,.
\end{equation}
Because of the continuity of $\xi(k_1)$, we know that $\xi(k_{01} + \delta) = \xi(k_{01}-\delta)$ when $\delta\rightarrow 0$. Then by adding the two equations together, we have
\begin{equation}
  \xi(k_1')-\xi(k_1) = \int_{\Omega'_{k_1,k_1'}}d^2k \,f_{xy}\,,
\end{equation}
in which $\Omega'_{k_1,k_1'}$ stands for the same the rectangular region as $\Omega_{k_1,k_1'}$, with Dirac points removed. This is consistent with the definition of $f_{xy}$, because it is well defined only when $\mathbf{k}$ is not on any Dirac points. Now we can write down the Wilson loop winding number in terms of the integral of $f_{xy}$:
\begin{equation}
  e_2 = \frac{1}{2\pi}\int_{\rm BZ'}d^2k\,f_{xy}\,,
\end{equation}
where BZ' is the whole Brillouin zone with Dirac points removed from the integration area.
\begin{figure}[!htbp]
\centering
\includegraphics[width=10cm]{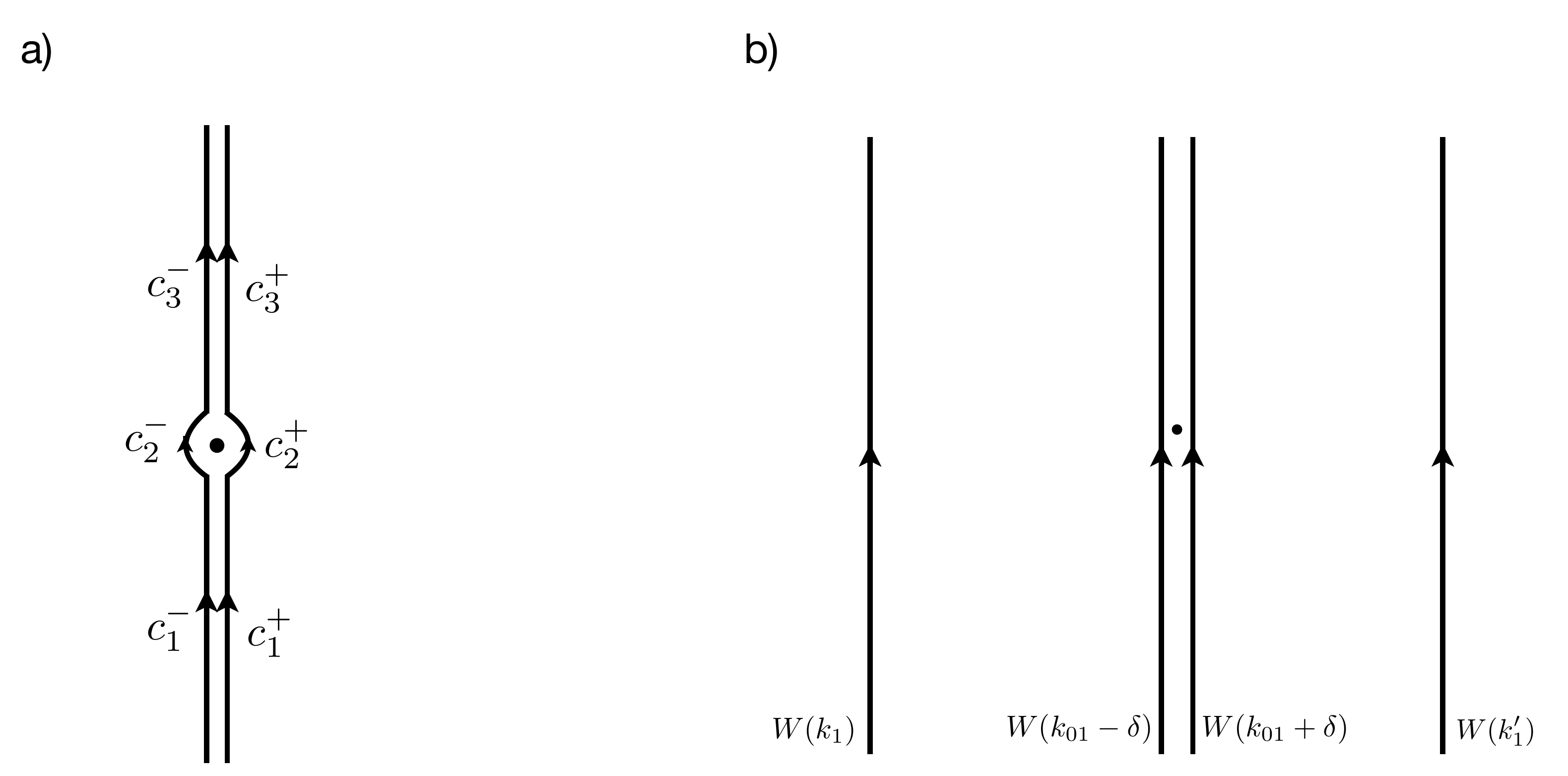}
\caption{a) The winding of $\mathbf{a}(\mathbf{k})$ around a Dirac point and it leads to a discontinuous $\int_0^{2\pi}dk_2\,a_2(k_x,k_y)$. b) The increment of Wilson loop eigenvalue between $k_1$ and $k_1'$ is given by the integral of $f_{xy}$ in the rectangular region between $k_1$ and $k_1'$ with Dirac points removed.}
\label{fig:windingstokes}
\end{figure}

\subsubsection{Fubini-Study metric of \texorpdfstring{$C_{2z}T$}{C2T} topology}
The $C_{2z}T$ symmetry yields a constraint on the quantum geometric tensor. The quantum geometric tensor satisfies the following equation:
\begin{equation}
  \mathfrak{G}_{ij}(\mathbf{k}) = B(\mathbf{k})\mathfrak{G}^*_{ij}(\mathbf{k})B^\dagger(\mathbf{k})\,,
\end{equation}
where $B(\mathbf{k})$ is the sewing matrix of the $C_{2z}T$ symmetry.  Similar to the Berry connection case, this equation gives a constraint on the matrix elements of the quantum geometric tensor. It can be shown that $\mathfrak{G}_{ij}$ can only take the following form:
\begin{equation}
  \mathfrak{G}_{ij} = \left( \begin{array}{cc}
    \mathfrak{g}^{11}_{ij} & \left(\gamma_{ij} +\frac12 f_{ij}\right)e^{ \frac{i}{2}(\theta_1 - \theta_2)}\\
    \left(\gamma_{ij} -\frac12 f_{ij}\right)e^{ \frac{i}{2}(\theta_2 - \theta_1)} & \mathfrak{g}^{22}_{ij}
    \end{array}\right)\,,
\end{equation}
in which $\gamma_{ij} = \mathfrak{g}^{12}_{ij}e^{\frac{i}{2}(\theta_2-\theta_1)}$ is a real symmetric tensor. By definition the Fubini-Study metric of the two flat bands with spin $\uparrow$ and valley $K$ in TBLG is given by $g_{ij} = \mathfrak{g}^{11}_{ij} + \mathfrak{g}^{22}_{ij}$. Similar to the Chern insulator case, if we choose $c_x = (1, i e^{\frac{i}{2}(\theta_2 -\theta_1)})^{\rm T}$, $c_y = (i, -e^{\frac{i}{2}(\theta_2 - \theta_1)})^{\rm T}$, then Eq. (\ref{eqn:provepositive}) can tell us
\begin{equation}
  \sum_{ij}c^\dagger_i \mathfrak{G}_{ij}c_j = \mathfrak{g}^{11}_{xx} + \mathfrak{g}^{22}_{xx} + \mathfrak{g}^{11}_{yy} + \mathfrak{g}^{22}_{yy} - 2f_{xy} = {\rm tr}\,{g} - 2f_{xy}\geq 0\,,~~~{\rm tr}\, {g} \geq 2f_{xy}\,,
\end{equation}
similarly, if we choose $c_x = (1, -i e^{\frac{i}{2}(\theta_2 -\theta_1)})^{\rm T}$, $c_y = (i, e^{\frac{i}{2}(\theta_2 - \theta_1)})^{\rm T}$, we will have ${\rm tr}\,g\geq -2f_{xy}$. In conclusion, we have ${\rm tr}\, g\geq 2|f_{xy}|$. As we have shown in the last subsection, the integral of $f_{xy}$ gives us the $C_{2z} T$ winding number $e_2$. We hence find that the integral of metric is bounded by the winding number:
\begin{equation}\label{eqn:c2tbound}
  \frac{1}{4\pi} \int_{\rm BZ} d^2k \,{\rm tr}\,g(\mathbf{k}) \geq \frac{1}{2\pi}\int_{\rm BZ'} d^2k\,|f_{xy}| \geq\Bigg{|}\frac{1}{2\pi}\int_{\rm BZ'} d^2 k\,f_{xy}\Bigg{|} = |e_2|\,.
\end{equation}

\section{Superconductivity in TBLG}\label{sec:sctblg}
We now discuss the superconductivity in TBLG. The TBLG system has two spin and two valley quantum number. Thus there are 8 flat bands in total. We denote the fermion annihilation operator by $\psi_{\eta,s,\mathbf{Q},a,\mathbf{k}}$, where $\eta$ stands for graphene valley, $s$ stands for electron spin, $\mathbf{Q}$ stands for the hexagonal lattice site in momentum space formed by $\mathbf{q}_{1,2,3}$ vectors and $a$ stands for graphene sublattice. The BM continuum model in Eq. (\ref{eqn:moireband}) is the Hamiltonian of spin $\uparrow$ and valley $K$. To obtain the Hamiltonian of valley $K'$, we can use the time reversal transformation $\mathcal{T}$. The fermion operators $\psi$ transform as shown under $\mathcal{T}$:
\begin{equation}
  \mathcal{T}\psi_{\eta,s,\mathbf{Q},a,\mathbf{k}}\mathcal{T}^{-1} =i(s_y)_{ss'} (\tau_x)_{\eta\eta'}V_{\mathbf{Q},\mathbf{Q}'} \psi_{v',s',\mathbf{Q}',a,-\mathbf{k}}\,,\label{eqn:sftrs}\\
\end{equation}
in which $s_i$, $\tau_i$ and $\sigma_i$ are Pauli matrices acting on spin, graphene valley and graphene sublattice indices, respectively. The matrix $V$ acting on $\mathbf{Q}$ index is defined by $V_{\mathbf{Q},\mathbf{Q}'} = \delta_{\mathbf{Q},-\mathbf{Q}'}$. Therefore the representation matrix of spinful time reversal transformation in TBLG is $U_T = is_y\tau_xV$.

By taking all the spin and valley into consideration, the kinetic energy $\mathcal{H}(\mathbf{k})$ and time reversal transformation representation $U_T$ can be written as the following diagonal block matrices:
\begin{equation}
  \mathcal{H}(\mathbf{k}) = \left(\begin{array}{cccc}
      h_{K\uparrow}(\mathbf{k}) &&&\\ &h_{K'\uparrow}(\mathbf{k})&& \\ && h_{K\downarrow}(\mathbf{k}) & \\ &&& h_{K'\downarrow}(\mathbf{k})
    \end{array}
  \right)\,,~~~U_T = is_y\tau_xV = \left(
    \begin{array}{cccc}
      &&&V \\ && V & \\ & -V && \\-V&&&
    \end{array}
  \right)\,,
\end{equation}
in which the entries of this matrix stand for $(K\uparrow), (K'\uparrow), (K\downarrow)$ and $(K'\downarrow)$. Because there is no spin-orbit-coupling in graphene, we have $h_{K\uparrow}(\mathbf{k}) = h_{K\downarrow}(\mathbf{k}) = h(\mathbf{k})$, where $h(\mathbf{k})$ is the BM continuum model in Eq. (\ref{eqn:moireband}), and $h_{K'\uparrow}(\mathbf{k}) = h_{K'\downarrow}(\mathbf{k}) = h'(\mathbf{k})$. Since the Hamiltonian $\mathcal{H}(\mathbf{k})$ is time reversal invariant, we have $U^\dagger_T\mathcal{H}^*(-\mathbf{k})U_T = \mathcal{H}(\mathbf{k})$. This condition tells us that $h'(\mathbf{k}) = Vh^*(-\mathbf{k})V$. Suppose the eigenvectors of $h(\mathbf{k})$ are denoted by $u_n(\mathbf{k})$, the flat bands wave function will be
\begin{equation}\label{eqn:tildeU}
  \tilde{U}_\mathbf{k} = \left(\begin{array}{cccc}
    \tilde{u}_\mathbf{k} &&& \\ & V\tilde{u}^*_{-\mathbf{k}} && \\ && \tilde{u}_\mathbf{k} & \\ &&& V\tilde{u}^*_{-\mathbf{k}}
  \end{array}
  \right)\,,
\end{equation}
where $\tilde{u}_{\mathbf{k}} = (u_1(\mathbf{k}),u_2(\mathbf{k}))$ and $u_{1,2}(\mathbf{k})$ are the eigenvectors of $h(\mathbf{k})$ which corresponds to the two flat bands. 

We assume that the pairing order parameter has the form of Eq. (\ref{eqn:deltaansatz}). This pairing order parameter means that the pairing happens between electrons which has opposite spin, valley and momentum, because the time reversal transformation $U_T = is_y\tau_x V$ can flip the spin and valley at the same time. Therefore, by using the result in Sec. \ref{sec:fbsw}, the superfluid weight of TBLG is given by Eq. (\ref{eqn:swflat}). However, since $\tilde{U}_\mathbf{k}$ are the eigenvectors which corresponds to the all 8 flat bands with all spins and valleys, we cannot directly use the lower bound of Fubini-Study metric of one spin one valley $g_{ij}(\mathbf{k})$. In fact, integrand in Eq. (\ref{eqn:swflat}) is related with $g_{ij}(\mathbf{k})$ as shown: 
\begin{align}
  &{\rm Tr}\left[\frac{1}{2}\left(\partial_{k_i}\tilde{U}^\dagger_\mathbf{k}\partial_{k_j}\tilde{U}_\mathbf{k} +\partial_{k_j}\tilde{U}^\dagger_\mathbf{k}\partial_{k_i}\tilde{U}_\mathbf{k}\right) +  \left( \tilde{U}^\dagger_\mathbf{k}\partial_{k_i}\tilde{U}_\mathbf{k}\tilde{U}^\dagger_\mathbf{k}\partial_{k_j}\tilde{U}_\mathbf{k}\right)\right]\nonumber\\
  =&2{\rm Tr}\left[\frac{1}{2}\left(\partial_{k_i}\tilde{u}^\dagger_\mathbf{k}\partial_{k_j}\tilde{u}_\mathbf{k} + \partial_{k_j}\tilde{u}^\dagger_\mathbf{k}\partial_{k_i}\tilde{u}_\mathbf{k}\right) +  \left( \tilde{u}^\dagger_\mathbf{k}\partial_{k_i}\tilde{u}_\mathbf{k}\tilde{u}^\dagger_\mathbf{k}\partial_{k_j}\tilde{u}_\mathbf{k}\right)\right] \nonumber\\ 
  &+ 2{\rm Tr}\left[\frac{1}{2}\left(\partial_{k_i}\tilde{u}^{\rm T}_{-\mathbf{k}}\partial_{k_j}\tilde{u}^*_{-\mathbf{k}} +\partial_{k_j}\tilde{u}^{\rm T}_{-\mathbf{k}}\partial_{k_i}\tilde{u}^*_{-\mathbf{k}}\right) +  \left(\tilde{u}^{\rm T}_{-\mathbf{k}}\partial_{k_i}\tilde{u}^*_{-\mathbf{k}}\tilde{u}^{\rm T}_{-\mathbf{k}}\partial_{k_j}\tilde{u}^*_{-\mathbf{k}}\right)\right]\nonumber\\
  =&2g_{ij}(\mathbf{k}) + 2 g^*_{ij}(-\mathbf{k})\,.
\end{align}
However, as we have mentioned, Fubini-Study metric defines a positive distance in the BZ, which means it must be a real tensor, hence $g^*_{ij}(-\mathbf{k}) = g_{ij}(\mathbf{k})$. Therefore we have
\begin{align}
   2\int\frac{d^2k}{(2\pi)^2}\left( g_{ij}(\mathbf{k}) + g^*_{ij}(-\mathbf{k}) \right)&= 2\int\frac{d^2k}{(2\pi)^2}\left( g_{ij}(\mathbf{k}) + g_{ij}(-\mathbf{k}) \right)\nonumber\\
  &= 2\int\frac{d^2k}{(2\pi)^2}g_{ij}(\mathbf{k}) + 2 \int\frac{d^2k}{(2\pi)^2}g_{ij}(-\mathbf{k})\nonumber\\
  &= 4\int\frac{d^2k}{(2\pi)^2}g_{ij}(\mathbf{k})\,.
\end{align}
Then by using Eq. (\ref{eqn:c2tbound}) the lower bound of this integral will be
\begin{equation}
  {\rm tr}\left\{\int\frac{d^2k}{(2\pi)^2}{\rm Tr}\left[\frac{1}{2}\left(\partial_{k_i}\tilde{U}^\dagger_\mathbf{k}\partial_{k_j}\tilde{U}_\mathbf{k} +\partial_{k_j}\tilde{U}^\dagger_\mathbf{k}\partial_{k_i}\tilde{U}_\mathbf{k}\right) +  \left( \tilde{U}^\dagger_\mathbf{k}\partial_{k_i}\tilde{U}_\mathbf{k}\tilde{U}^\dagger_\mathbf{k}\partial_{k_j}\tilde{U}_\mathbf{k}\right)\right]\right\} \geq \frac{4}{\pi}\,.
\end{equation}
That naturally leads to the lower bound of the trace of the superfluid weight: ${\rm tr}\,D_s \geq \frac{8(\Delta_1 + \Delta_2)}{\pi}\sqrt{\nu(1-\nu)}$. Here we used the fact that the winding number of the flat bands in TBLG is one \cite{songz2018}. Due to $C_{3z}$ symmetry, $D_s$ must proportional to the identity, therefore we obtain $D_s \geq \frac{4(\Delta_1+\Delta_2)}{\pi}\sqrt{\nu(1-\nu)}$.

\section{Finite temperature study and Berezinskii-Kosterlitz-Thouless transition}\label{sec:ftbkt}
As mentioned in the main text, we will assume $\Delta_2 = 0$. Before we estimate the actual value of topological lower bound, we need to know the gap function $\Delta_1$ at zero temperature. As we discussed in the main text, the transition temperature measured in experiment is the Berezinskii-Kosterlitz-Thouless temperature $T_c$, and it is different from the mean field temperature $T_c^*$ where Cooper pairing disappears. The condition of BKT transition is universal and it is given by
\begin{equation}\label{eqn:ktcondition}
  \frac{D_s(T_c)}{T_c} = \frac{8}{\pi}\,, 
\end{equation} 
where $D_s(T)$ is the superfluid weight at temperature $T$. When this transition happens, the superfluid weight jumps from a finite value to zero, which is totally different from the situation in 3D. Hence the BKT transition temperature is always lower $T_c < T_c^*$. Now we use the mean field Hamiltonian to determine the ratio of BKT temperature and mean field temperature. The free energy at finite temperature is given by the following equation:
\begin{equation}
  \Omega(T,\mathbf{q}) = -T\sum_{\mathbf{k},n}\log\left[2 \cosh\left(\frac{E_{\mathbf{k}n}(\mathbf{q})}{2T}\right)\right]\,,
\end{equation}
in which $E_{\mathbf{k}n}(\mathbf{q})$ are the \emph{positive} eigenvalues of $\tilde{\mathscr{H}}_\mathbf{k}(\mathbf{q})$. Now we take the derivative of this free energy, and the superfluid weight will be
\begin{equation}\label{eqn:freeenergyT}
  \frac{1}{V}\frac{\partial^2\Omega(T,\mathbf{q})}{\partial{q_i}\partial{q_j}}\Big{|}_{\mathbf{q}=0}= -\frac{1}{2}\int\frac{d^2k}{(2\pi)^2}\sum_n\left[\frac{\partial_{q_i}E_{\mathbf{k}n}(\mathbf{q})\partial_{q_j}E_{\mathbf{k}n}(\mathbf{q})}{2T\cosh^2\left(\frac{E_{\mathbf{k}n}(\mathbf{q})}{2T}\right)}\Bigg{|}_{\mathbf{q}=0} + \partial_{q_i}\partial_{q_j}E_{\mathbf{k}n}(\mathbf{q})\tanh\left(\frac{E_{\mathbf{k}n}(\mathbf{q})}{2T}\right)\Bigg{|}_{\mathbf{q}=0}\right]\,.
\end{equation}
In Sec. \ref{sec:fbsw} we have shown that $\partial_\mathbf{q}\lambda_{\mathbf{k}n}(\mathbf{q})|_{\mathbf{q}=0} = \partial_\mathbf{q}\varphi_{\mathbf{k}n}(\mathbf{q})|_{\mathbf{q}=0} = 0$, so all the first order derivative terms in Eq. (\ref{eqn:freeenergyT}) will disappear. Also in flat band limit, $E_{n\mathbf{k}}(0)$ is independent with $\mathbf{k}$ and $n$ under the pairing ansatz Eq. (\ref{eqn:deltaansatz}), thus the superfluid weight at finite temperature will become
\begin{equation}
  [D_s]_{ij}(T) = -\frac12\tanh\left(\frac{\sqrt{(\varepsilon_0 - \mu)^2 + \Delta_1^2}}{2T}\right) \sum_n\int\frac{d^2k}{(2\pi)^2} \partial_{q_i}\partial_{q_j}E_{\mathbf{k}n}(\mathbf{q})\,,
\end{equation}
and we find this result only differs by a hyperbolic tangent factor from the expression at zero temperature in Sec. \ref{sec:fbsw}. By using Eq. (\ref{eqn:flatbandsw}), we can write down the expression of $D_s$ at finite temperature. Since it is isotropic due to $C_{3z}$ symmetry, the superfluid weight can be written as
\begin{equation}\label{eqn:Tsw}
  D_s(T) = \frac{\Delta_1(T)^2}{2\sqrt{(\varepsilon_0-\mu)^2 + \Delta_1(T)^2}}\tanh\left(\frac{\sqrt{(\varepsilon_0-\mu)^2 + \Delta_1(T)^2}}{2T}\right)\int \frac{d^2k}{(2\pi)^2}{\rm tr\,}g(\mathbf{k})\,,
\end{equation}
here $\Delta_1(T)$ is the gap function at temperature $T$, and $g(\mathbf{k})$ is the Fubini-Study metric derived from \emph{all} the flat bands. For convenience we denote $\sqrt{(\varepsilon_0-\mu)^2 + \Delta_1(T)}$ by $E_B(T)$. Notice that this expression depends on the chemical potential $\mu$ instead of the filling ratio $\nu$. The particle number is no longer simply given by the Bogoliubov quasiparticle coefficient $|v_\mathbf{k}|^2$ here because of thermal excitations. Because both the free fermion band and Bogoliubov quasiparticle spectrum $E_B(T)$ are momentum independent, the filling ratio of flat bands are given by
\begin{equation}
  \nu = \langle c^\dagger_{n\mathbf{k},K\uparrow}c_{n\mathbf{k},K\uparrow}\rangle = |u|^2\langle \gamma^\dagger_{\mathbf{k},K\uparrow}\gamma_{\mathbf{k},K\uparrow} \rangle + |v|^2 \langle \gamma_{-\mathbf{k},K'\downarrow}\gamma^\dagger_{-\mathbf{k},K'\downarrow} \rangle\,,
\end{equation}
here $u^2,v^2 = \frac12 \left(1\pm\frac{\varepsilon_0-\mu}{E_B(T)}\right)$ are the coefficients of Bogoliubov transformation, $c,c^\dagger$ are the electron operators and $\gamma,\gamma^\dagger$ are the Bogoliubov quasiparticle operators. When $T\neq 0$, the expectation value of Bogoliubov quasiparticle number is given by Fermi-Dirac distribution, therefore we have $\langle \gamma^\dagger_{\mathbf{k},K\uparrow}\gamma_{\mathbf{k},K\uparrow} \rangle = n_F(E_B(T))$ and $\langle \gamma_{-\mathbf{k},K'\downarrow}\gamma^\dagger_{-\mathbf{k},K'\downarrow} \rangle = 1 +n_F(E_B(T))$. Hence the filling ratio will be 
\begin{equation}\label{eqn:particlenumber}
   \nu = v^2 + n_F(E_B(T)) = \frac12\left(1-\frac{\varepsilon_0-\mu}{E_B(T)}\right)+\frac{1}{e^{\beta E_B(T)}+1}\,.
\end{equation}
If the temperature $T$, the order parameter $\Delta_1(T)$ and the filling ratio $\nu$ are given, the chemical potential and further the Bogoliubov energy $E_B(T)$ can be solved from Eq. (\ref{eqn:particlenumber}). Finally we are able to use Eq. (\ref{eqn:Tsw}) to obtain the superfluid weight with given temperature $T$ and filling ratio $\nu$.

To determine the relationship between BKT temperature and mean field temperature, we have to know the value of $D_s(T)$. Here we use flat bands of TBLG as an example. For simplicity, we use the lower bound of the metric to estimate its integral:$\int \frac{d^2k}{(2\pi)^2} {\rm tr\,}g \approx\frac{4}{\pi}$. Hence the superfluid weight becomes
\begin{equation}
  D_s(T) \approx \frac{2\Delta_1^2(T)}{\pi\sqrt{(\varepsilon_0-\mu)^2 + \Delta_1^2(T)}}\tanh\left(\frac{\sqrt{(\varepsilon_0-\mu)^2 + \Delta_1^2(T)}}{2T}\right)\,,\label{eqn:valuesw}
\end{equation}
in which the chemical potential is related with filling ratio $\nu$ by Eq. (\ref{eqn:particlenumber}). Since it is an order of magnitude estimation here, we can simply use the BCS mean field result $\Delta_1(T) \approx \Delta_1(0)\left(1 - \frac{2T}{\Delta_1(0)}\right)^{1/2}$ to determine the temperature dependent superconducting gap. Then we can numerically evaluate the value of Eq. (\ref{eqn:valuesw}), and the result is shown in FIG. \ref{fig:KT} when $\nu = 1/4$. The intersection of the solid line and the dotted line is the KT transition point, which satisfies Eq. (\ref{eqn:ktcondition}). We find that when $\nu = 1/4$, the transition temperature is $T_c \approx 0.34 T_c^* \approx 0.17 \Delta_1(0)$. Here we were using the lower bound of superfluid weight, which means the ratio $T_c/T_c^*$ is under estimated.

\section{The validity of BCS wave function in the flat band limit}\label{sec:validity}
By using Hund's rule we argue that the BCS mean field ground state is a good approximation of the real ground state. This argument was also applied in the spin Chern number case \cite{Peotta2015} and originally this idea comes from quantum Hall ferromagnet \cite{spielman2000quantumhallferro}. The BCS wave function can be written as
\begin{equation}
  |{\rm BCS}\rangle = \prod_{\mathbf{k},n}\left(\sqrt{1-\nu} + \sqrt{\nu}c^\dagger_{K\uparrow,\mathbf{k}n}c^\dagger_{K'\downarrow,-\mathbf{k}n}\right)\left(\sqrt{1-\nu} + \sqrt{\nu}c^\dagger_{K'\uparrow,\mathbf{k}n}c^\dagger_{K\downarrow,-\mathbf{k}n}\right)|0\rangle\,,
\end{equation}
in which $n$ is the flat band index. This wave function is equivalent to the following form:
\begin{equation}
  |{\rm BCS}\rangle = \prod_{\mathbf{k},n}\left(\sqrt{1-\nu} c_{K'\downarrow,-\mathbf{k}n} + \sqrt{\nu}c^\dagger_{K\uparrow,\mathbf{k}n}\right)\left(\sqrt{1-\nu} c_{K\downarrow,-\mathbf{k}n} + \sqrt{\nu}c^\dagger_{K'\uparrow,\mathbf{k}n}\right)c^\dagger_{K'\downarrow,-\mathbf{k}n}c^\dagger_{K\downarrow,-\mathbf{k}n}|0\rangle\,,
\end{equation} 
If we perform the following particle-hole transformation: $c^\dagger_{\downarrow,\mathbf{k}} \rightarrow c_{\downarrow,-\mathbf{k}}$. That transformation changes the attractive interaction into repulsive. Originally, the interaction between electrons with opposite spins is attractive. After the transformation, it becomes an attractive interaction between spin up electrons and spin down holes, which is equivalent to a repulsive interaction between electrons with opposite spin. In order to show this clearly, we use the following toy model Hamiltonian  
\begin{equation}
  H_{\rm int} = \sum_{v,n,\mathbf{k},\mathbf{k}',\mathbf{q}}  g(\mathbf{q}) c^\dagger_{v,\uparrow,n,\mathbf{k} + \mathbf{q}} c^\dagger_{\bar{v},\downarrow,n,\mathbf{k}'-\mathbf{q}} c_{\bar{v},\downarrow,n,\mathbf{k}'} c_{v,\uparrow,n,\mathbf{k}}\,,
\end{equation}
where $g(\mathbf{q})$ is the Fourier transformation of an attractive potential, and $v$ stand for valley, and $\bar{v}$ is the opposite valley of $v$. After this particle-hole transformation, the toy model Hamiltonian becomes the following form:
\begin{equation}
  H_{\rm int} \rightarrow -\sum_{v,n,\mathbf{k},\mathbf{k}',\mathbf{q}} g(\mathbf{q}) c^\dagger_{v,\uparrow,n,\mathbf{k}+\mathbf{q}} c^\dagger_{\bar{v},\downarrow,n,-\mathbf{k}'}c_{\bar{v},\downarrow,n,-\mathbf{k}'+\mathbf{q}} c_{v,\uparrow,n,\mathbf{k}}\,.
\end{equation}
We notice that the sign of $g(\mathbf{q})$ is flipped, which means a repulsive interaction. However, by switching the creation and annihilation operators, the vacuum state is also changed: $\prod_{\mathbf{k},n}c^\dagger_{K'\downarrow,-\mathbf{k}n}c^\dagger_{K\downarrow,-\mathbf{k}n}|0\rangle \rightarrow |0\rangle $. Therefore, under this particle-hole transformation, the BCS wave function becomes
\begin{equation}
  |{\rm BCS}\rangle \rightarrow \prod_{\mathbf{k},n}\left(\sqrt{1-\nu} c^\dagger_{K'\downarrow,\mathbf{k}n} + \sqrt{\nu}c^\dagger_{K\uparrow,\mathbf{k}n}\right)\left(\sqrt{1-\nu} c_{K\downarrow,\mathbf{k}n}^\dagger + \sqrt{\nu}c^\dagger_{K'\uparrow,\mathbf{k}n}\right)|0\rangle\,.
\end{equation}
If we use $(K\uparrow)$ and $(K'\downarrow)$ or $(K\downarrow)$ and $(K'\uparrow)$ to form a pseudo spin $\frac12$ space, then $|\rm BCS\rangle$ transforms into a ferromagnetic state of the pseudo spin. Since the ferromagnetic state is symmetric in the pseudo spin space, its orbital wave function is totally antisymmetric. Therefore, this ferromagnetic state is energetically favored by repulsive interaction, and we conclude that BCS wave function can be a good approximation to the ground state.

\end{widetext}

\end{document}